\documentclass[%
 reprint,
 amsmath,amssymb,
 aps,
]{revtex4-2}

\usepackage{graphicx}
\usepackage{dcolumn}
\usepackage{bm}

\usepackage{graphicx}
\usepackage{dcolumn}
\usepackage{bm}
\usepackage{xcolor}
\usepackage{ulem}	
\usepackage{amssymb}  
\usepackage{amsmath}  

\begin{document}
\preprint{APS/123-QED}

\title{
Barkhausen noise in disordered strip-like ferromagnets: experiment vs. simulations
}

\author{Djordje Spasojevi\'c}
\affiliation{Faculty of Physics, University of Belgrade, POB 44, 11001 Belgrade, Serbia}
\author{Milo\v s Marinkovi\'c}
\affiliation{Faculty of Physics, University of Belgrade, POB 44, 11001 Belgrade, Serbia}
\author{Dragutin Jovkovi\'{c}}
\affiliation{Faculty of Mining and Geology, University of Belgrade, POB 162, 11000 Belgrade, Serbia}
\author{Sanja Jani\'cevi\'c}
\affiliation{Faculty of Science, University of Kragujevac, POB 60, 34000 Kragujevac, Serbia}
\author{Lasse Laurson}
\affiliation{Computational Physics Laboratory, Tampere University, POB 692, FI-33014 Tampere, Finland} 
\author{Antonije Djordjevi\'{c}}
\affiliation{School of Electrical Engineering, University of Belgrade, 11000 Belgrade, Serbia}
\affiliation{Serbian Academy of Sciences and Arts, 11000 Belgrade, Serbia}

\begin{abstract}
In this paper, we compare the results obtained from the low-frequency Barkhausen noise recordings performed on a VITROPERM 800 metallic glass sample with the results from the numerical simulations of the Random Field Ising Model systems. We show that when the values of disorder and driving rates of the model systems are suitably chosen, a considerable matching with the experimental results is achieved indicating that the model can reproduce a majority of the Barkhausen noise features.
\end{abstract}

\maketitle


\section{Introduction}\label{intro}

A multitude of systems respond to slowly changing external conditions by exhibiting a bursty, "crackling noise" type of response, consisting of a sequence of bursts of activity, or avalanches, characterized by scale-free power-law distributions~\cite{CracklingNoise}. A prime example of both fundamental and applied interest is given by
 Barkhausen noise (BN)~\cite{Barkhausen}. It consists of irregular electromotive force (EMF) pulses induced by the jumps in magnetization caused by the jerky motion of magnetic domain walls in response to the slow changes of the external magnetic field. 
 
 The BN studies performed so far \cite{ABBM,Spasojevic1996,Bahiana,Mehta,Colaiori,DZ2000,Puppin,Moore,Puppin2,Kim,Ryu,Shin,Lee,Bohn1,Bohn2,DurPRL2016,Bohn3,LasseTEM} have shown that the (individual) jumps, and therefore the BN pulses, are stochastic and that their distributions follow  power laws described by power-law indices which satisfy certain scaling relations. 
 Experimental studies of Barkhausen noise have been conducted both in bulk \cite{DZ2000} and nonequilateral geometry samples, in particular thin films \cite{Puppin,Moore,Puppin2}. Two distinctive universality classes have been identified in polycrystalline and amorphous bulk materials, while in thin samples the question of universality remains to be resolved.  
The use of magneto-optical techniques \cite{Ryu} demonstrated the clear difference between the magnetic behavior of thin films and bulk materials, while more scrutinized investigations \cite{Bohn1,Bohn2,Bohn3} confirmed the existence of various types of 2D dynamics in thin films with different thicknesses.
BN is by no means unique in exhibiting crackling noise with power law statistics. Numerous other systems exhibit a similar pattern of behavior sharing the profound analogies despite fundamental differences in spatio-temporal scales, system geometry, structure, underlying interactions and type of driving.
Examples of such systems are covering a broad range from compressed nanocrystals \cite{nano}, to imbibition fronts in porous materials \cite{imbib}, to plastic deformation due to collective dynamics of dislocations \cite{plastic,Ispanovity2014,Jstat2015,Ispanovity2022}, heartbeat \cite{heart} and brain dynamics \cite{brain,DahmenPRL2012,MillerSciRep2019}, earthquakes \cite{FISH98,Earthquakes,Petri} to financial stock markets \cite{financial,Financial2013}. 

Theoretical or numerical models of Barkhausen noise (BN) proposed in the literature include discrete spin models such as the Random Field Ising Model (RFIM) \cite{BelangerNatterman,SethnaPRL93,OlgaPRB1999,Sethna2006}, micromagnetic simulations~\cite{herranen,kaappa}, and various domain wall models describing domain walls as driven elastic interfaces in random media~\cite{stanley}. In this paper, we consider the non-equilibrium RFIM as a model system, and present a comparison of the simulation results with low-frequency BN recordings performed on a VITROPERM 800 metallic glass sample. We find that by tuning the disorder and driving rate parameters of the model appropriately, the model is able to reproduce most of the experimental BN features.
Unlike 
in the equilibrium model version \cite{FontViv,LiuDah,LiuDah2,Balog}, in 
which the system evolves jumping between the 
equilibrium states determined by the current 
value of the external magnetic field, in the 
nonequilibrium model version, the system evolves 
through the nonequilibrium states following some 
local dynamic rule. The nonequilibrium version enables 
simulations of the time response of the system 
(mimicking BN) under all driving 
conditions and better corresponds to the BN 
experiments performed on metallic glasses which 
are by themselves not in equilibrium. Also, this 
model version is greatly 
computationally more efficient, so 
it enables simulations of much larger systems 
reducing the finite-size effects.

This paper is organized as follows: after the introduction in Section \ref{intro}, the details on the experimental analysis are presented in Section \ref{ExpAn}, including Subsections \ref{ExpSetup} and \ref{BNRecs} devoted to the description of the experimental setup and BN recordings. Section \ref{results} showcases obtained results and contains Subsections \ref{ExpBNLoopsDecomps} and \ref{ExpBNDistribss} dedicated to the analysis of experimental BN response signals and hysteresis loops, a detailed explanation of the decomposition of BN signal into BN pulses, and experimental BN distributions, followed by Subsection \ref{Comp} devoted to the comparison of results of experimental BN measurements
and numerical simulations of RFIM. The paper ends with discussion and conclusion presented in Section \ref{concl}.

\section{EXPERIMENTAL ANALYSIS}\label{ExpAn}

\subsection{Experimental setup}\label{ExpSetup}

\begin{figure}[hptb!]
\begin{center}
\includegraphics[width=8.5cm,trim=0cm 0cm 0cm 0cm,clip=true]{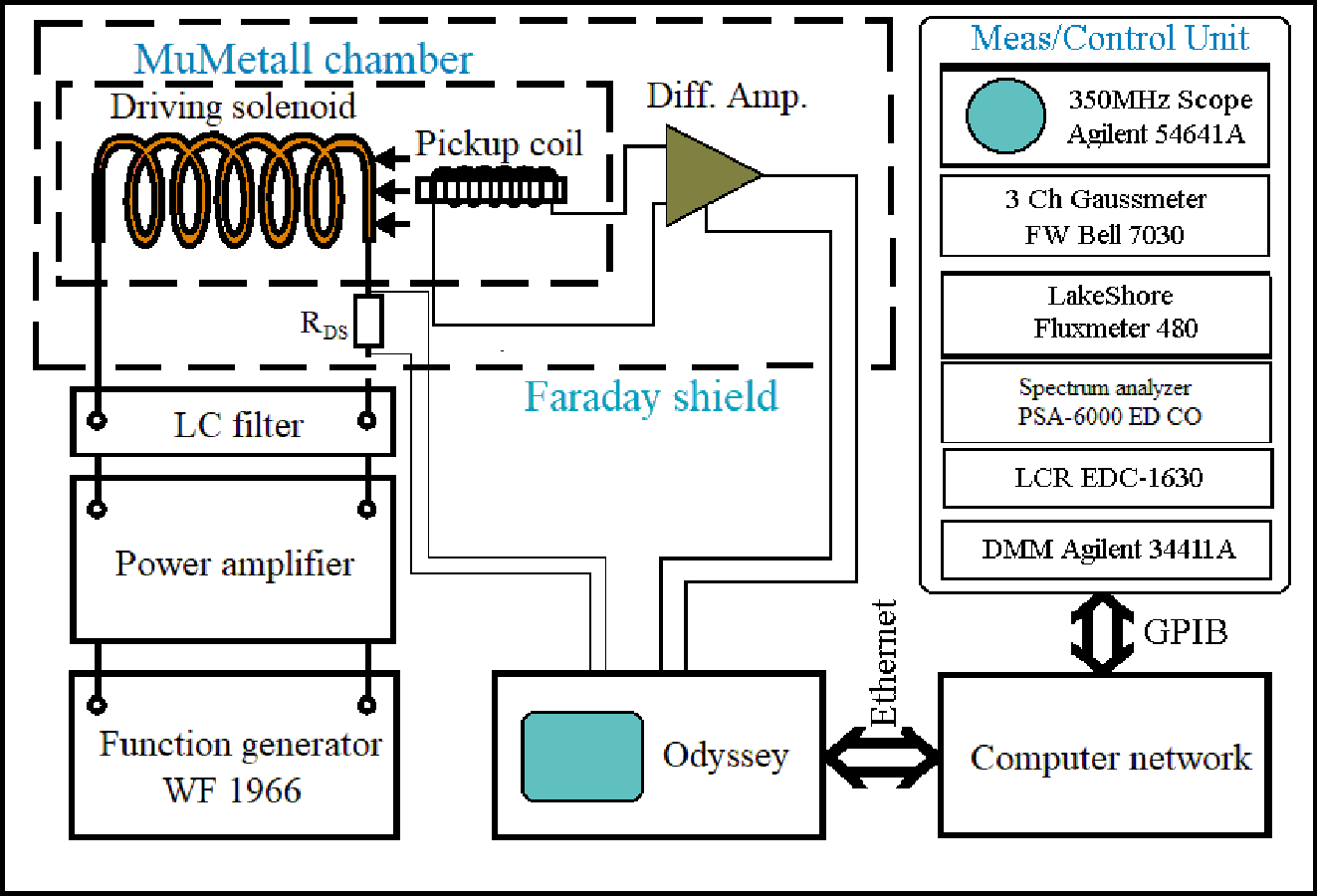}
\caption{
Schematic presentation of the experimental setup used in our Barkhausen noise measurements.
}
\label{Fig1}
\end{center}
\end{figure}

We performed the Barkhausen noise recordings using the setup depicted in Fig. \ref{Fig1}.
The Barkhausen noise signal was collected by a pickup coil  (1100 turns of AWG 34 copper 
wire 0.160~mm in diameter and 23.3~$\Omega$ resistance) tightly wound around the entire sample. The 
sample and the pickup coil were placed in the middle of a 30 cm long driving solenoid with a circular 
cross-section 5 cm in (inner) diameter, 840 turns in three rows of AWG 26 copper wire 0.405~mm in diameter and 5.7$~\Omega$ 
resistance. This driving coil provided the external magnetic field that is homogeneous along 80\% of its axis with 1\% tolerance due to 
the presence of two compensating coils (each with 50 turns of wire)  placed at the driving coil ends. 

The electric current, causing the external magnetic field by its flow through the 
driving coil, is produced  at the output of a dedicated battery-operated transconductance amplifier
from the input voltage signal generated by a function generator  (two-channel Multifunction synthesizer WF 1966B, NF corporation, Japan, 
0.01 $\mu$Hz - 50 MHz frequency range with $\pm 5$~ppm accuracy, 14-bit waveform resolution and $\pm 10$ V maximum output). 
The (digital high-frequency) generator voltage noise is suppressed at the amplifier input by 
a low-pass active filter (4 stages: Sallen-Key 20 Hz cutoff frequency, 12 dB/octave each, 48 dB/octave overall), 
so that the standard deviation of the generated current noise was  less than $20$ $\mu$A providing  driving  virtually without  
 digital and environmental noise.


The response signal 
(i.e. the electromotive force generated in 
the pickup coil) is routed from the pickup coil to a lab-made  
battery-operated  
two-stage voltage amplifier. In its first stage 
a CA-261F2 Low Noise Bipolar Amplifier (NF Corporation, Japan) is used with DC to 200 kHz frequency response, gain 40 dB $\pm$ 0.2 dB, and equivalent input noise voltage of 0.8 nV/$\sqrt{\mathrm{Hz}}$. In the second amplifier stage a National Semiconductor LME 49720NA Dual High Performance, High Fidelity Audio Operational Amplifier is used with maximum gain 
26.85~dB and equivalent input noise of 2.7 nV/$\sqrt{\mathrm{Hz}}$. 
The measurements were performed with the overall amplifier gain set at 2100,  turned on a first-order RC filter in the feedback section of the amplifier with the cutoff frequency set to 160 kHz, and the overall 
equivalent input noise voltage less than 1~nV/$\sqrt{\mathrm{Hz}}$. 

The amplified BN signal is led 
by a BNC cable to an OD200 acquisition 
card (four 
differential channels, 10 MHz/500 kHz maximum/continuous  sampling rate on all channels, and 14-bit/16-bit resolution) 
of the Odyssey XE (Nicolet, USA) data acquisition system. 
Besides the BN signal, we also monitored on another OD200  
channel the voltage at a $R_{\mathrm{DS}}=1.00\>\Omega$ (metal film) resistor 
connected in series with the driving solenoid giving us, via the conversion factor of $2960\> {\mathrm{Am}}^{-1}/{\mathrm{V}}$,
the time profile of the magnetic field $H$ inside the driving solenoid.

The sample, pickup coil, and driving solenoid were enclosed in 
a cylindrical MuMetall chamber (Vacuumschmelze, Germany) with 
four walls (each 3 mm thick) providing 50 cm high 
and 35 cm in diameter shielded volume, placed together with 
both amplifiers inside a sealed 1 m $\times$ 1 m $\times$ 1 m  
sound isolated Faraday shield  made of 1~cm thick solid aluminium. 
Due to such shielding and battery-operated amplifiers, the 
recorded Barkhausen noise was virtually free from the external 
electromagnetic noise and pollution penetrating from the 
electric network, as well as from the external static/low 
frequency environment electric and magnetic field. 

Let us also mention that in the calibration of the measurement system and for various types of control measurements  we used 350 MHz oscilloscope 
Agilent 54641A, Three-channel Gaussmeter FW Bell 7030, LakeShore Fluxmeter 480, 6.2 GHz Spectrum analyzer PSA-6000 (EdCo, Korea), 
LCR meter EDC-1630  (EdCo, Korea), and Digital multimeter Agilent 34411A, whereas for the control of the  ambiental electromagnetic noise and  distribution of static voltage
we used the TRIFIELD EMF Meter Model TF2 and Surface DC voltmeter SVM2 (AlphaLab, USA).

\subsection{Barkhausen noise recordings}\label{BNRecs}

In this paper, we present the results of the BN measurements performed on a 
16 cm $\times$ 1 cm $\times$ 40 $\mu{\mathrm{m}}$  sample of Vitroperm 800 supplied from Vacuumschmelze (VAC), Germany. 
Vitroperm 800 is a commercial nano-crystalline ferromagnetic alloy (82.8 Fe\%, 1.3\% Cu, 5.6\% Nb, 8.8\% Si, 1.5\% B by weight) 
with crystal grains of diameter between approx. 10-15 nm surrounded by an amorphous residual phase. It has 1.24 T saturation magnetic polarization 
reached at the field strength above approx. 500~A/m, very high treatment-dependent initial susceptibility (up to 600 000), negligible 
magnetostriction, and 600 ${}^\circ$C Curie temperature. This sample was annealed for 12 hours at 300 ${}^\circ$C without a magnetic field.

The Barkhausen noise recordings were performed in the vertically oriented external magnetic field $H$ parallel to the sample's longest side, which varied in time between -550 A/m and 550 A/m according 
to a periodic zero-mean triangle time 
profile at 0.5~mHz, 1~mHz, 2~mHz, 5~mHz, 
10~mHz, 20~mHz, and 50~mHz frequency. 
For each frequency (0.5-10~mHz)/(20-50~mHz), the presented data are measured in the 10 V/20 V voltage span
during 20 field cycles preceded by 50 cycles to reach 
a closed hysteresis loop.  The measurements were performed in the continuous acquisition mode of the Odyssey XE data acquisition system 
 at the 200 kS/s sampling rate with 16-bit resolution and turned on the internal anti-alias (low-pass) analog filter at 100~kHz cutoff frequency. The recorded data were stored in a 36 GB internal Odyssey XE acquisition 
 SCSI hard disk, and after completion of measurements transferred to a hard disk of an external PC connected to Odyssey 
 for the purpose of storage and numerical processing. 
  All measurements were performed during weekend nights to minimize the amount of external (environmental and electric network) noise.

\section{Results}\label{results}

\subsection{Experimental BN: response signal, hysteresis loops, and  decomposition of BN signal into BN pulses}\label{ExpBNLoopsDecomps}

\begin{figure}[hptb!]
\begin{center}
\includegraphics[width=8.5cm,trim=0cm 0.5cm 0cm 0.5cm,clip=true]{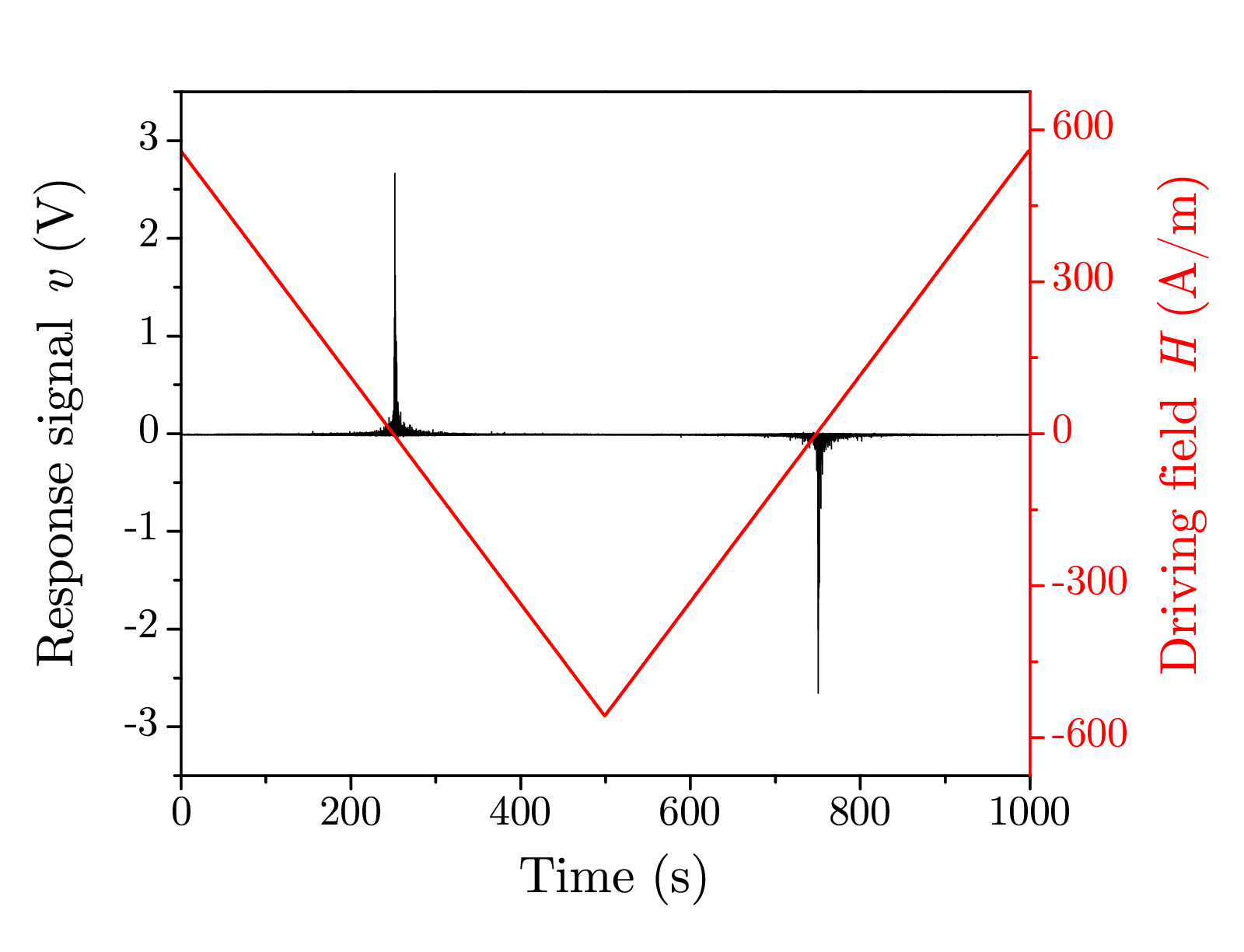}
\caption{
Time profiles  of the response voltage 
signal $v$ 
(black line) and the external magnetic 
field $H$ (red line) driving the sample 
at the 1~mHz frequency shown during one 
period. In our BN measurements, we connected 
the pickup coil so that the response 
signal had the polarity opposite to the 
sign of $dH/dt$. 
}
\label{Fig2}
\end{center}
\end{figure}

\begin{figure*}[hptb!]
\begin{center}
\includegraphics[width=\textwidth,trim=0cm 1.5cm 0cm 0cm,clip=true]{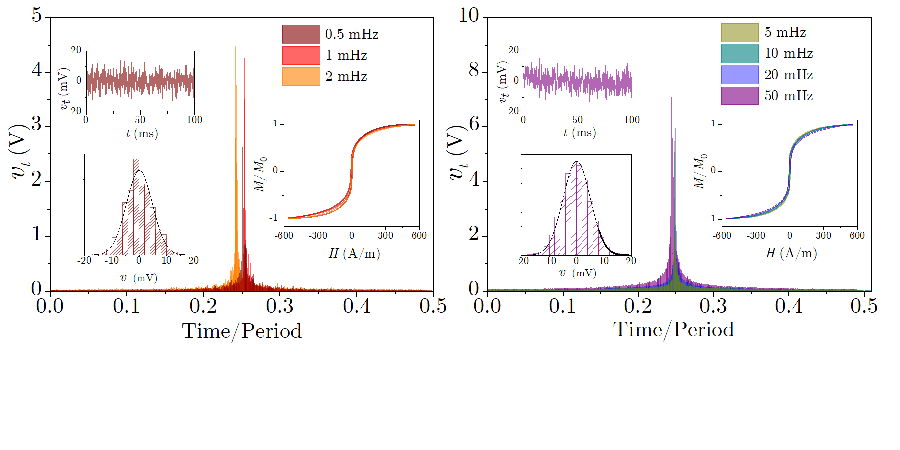}
\caption{
Main panels show one example of time 
profiles 
of the voltage response signal $v_t$ 
recorded during a single 
half-period 
of the external driving field $H$ 
for each of the 
employed "slow" (left) and "fast" 
(right) 
driving frequencies $\Omega$ quoted in legends. Top-left inset in the left main panel presents 
an excerpt of the time profile of the 
response signal recorded at 
$\Omega=0.5$~mHz near the maximum 
value of the external field $H$, while the  
histogram of presented values  
illustrates in the bottom-left inset of the same panel that 
these values are 
normally distributed, which could be mainly 
attributed to random fluctuations of 
sample's magnetization. In the right panel, 
the left insets show the same, but for 
$\Omega=50$~mHz, while for the remaining frequencies, the corresponding distributions are roughly the same with the standard deviation less than $5$~mV. 
Hysteresis curves, displaying 
versus the external magnetic field $H$ the sample's magnetization $M$ 
scaled by maximum magnetization $M_0$,  are 
given in the right insets.
}
\label{Fig3}
\end{center}
\end{figure*}

In Fig. \ref{Fig2} we illustrate the  time profiles of the response signal and the external magnetic field recorded at 1~mHz driving frequency during one period. 
 Examples of time profiles of response signals recorded in one half-period for each of the employed driving frequencies 
 are shown in the main panels of Fig. \ref{Fig3}, 
 for (0.5 - 2~mHz)/(5 - 50~mHz) on the left/right 
 main panel, together with the corresponding 
 hysteresis loops presented in the right insets. 
  Short excerpts of the response signal voltage, 
  recorded around the maximum value of $H$ when 
  the sample is saturated,  show in top-left 
  insets the overall noise of the measurement 
  system and the sample at $\Omega=0.5$~mHz (in 
  the left main panel) and $\Omega=50$~mHz (in 
  the right main panel). As illustrated by noise 
  histograms in the bottom-left insets, this 
  noise is of the Gaussian type with a standard 
  deviation 
  of less than 5~mV at these two and likewise for the remaining frequencies, so the Signal-to-Noise-Ratio by amplitude was maintained above 1000.

\begin{figure}[hptb!]
\begin{center}
\includegraphics[width=8.5cm,trim=0.5cm 0.5cm 0.5cm 1cm,clip=true]{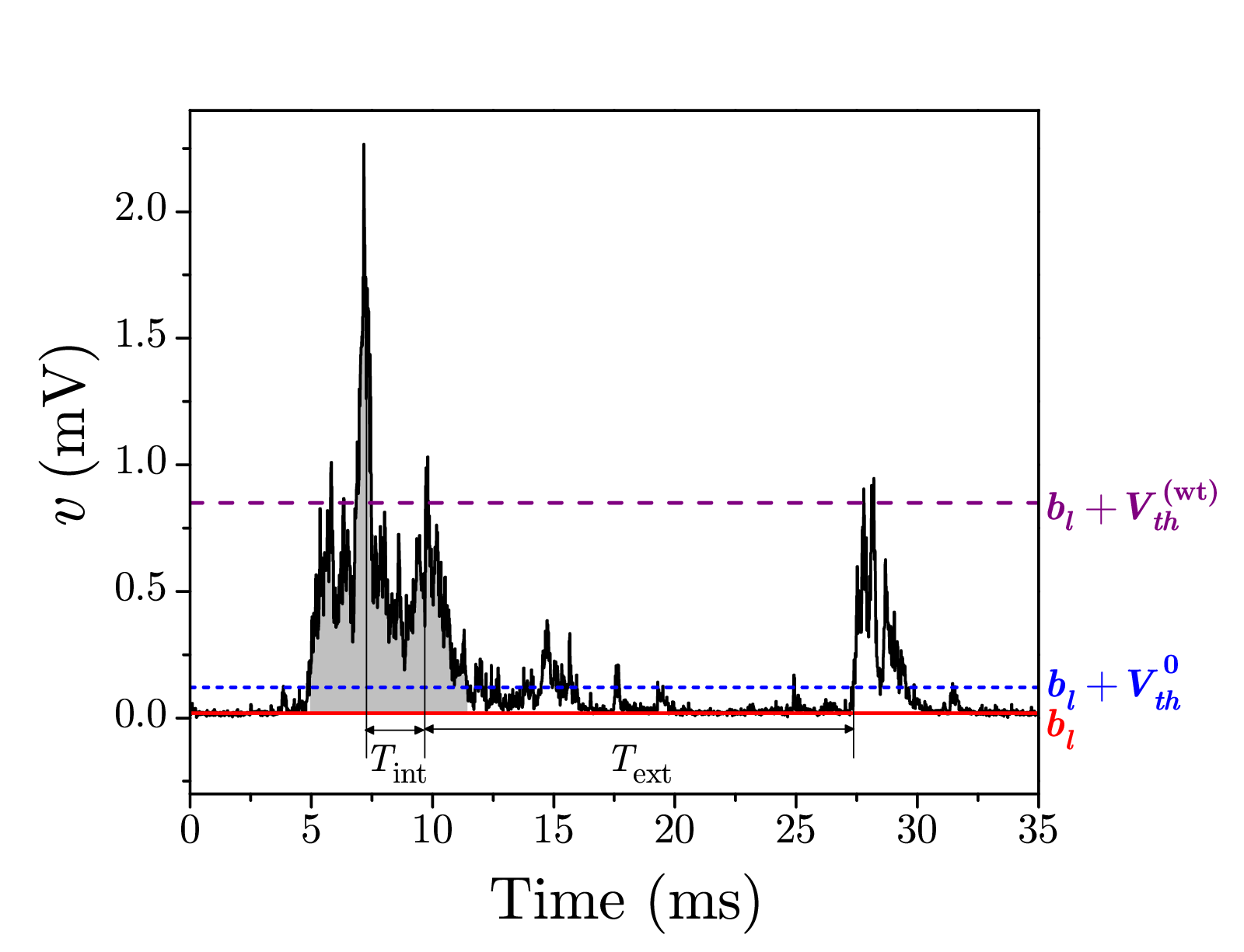}
\caption{ Part of the time-profile of 
 the response signal $v$ (black 
line) and horizontal lines representing (bottom to top): the baseline $b_l$ (red full line), base 
threshold level $b_l+V_{\mathrm{th}}^0$ (blue dotted line), and 
waiting-time 
threshold level $b_l+V_{\mathrm{th}}^{\mathrm{(wt)}}$ (purple dashed line). 
Base threshold $V_{\mathrm{th}}^0$ is used in the 
decomposition of the response signal into BN 
pulses (one such pulse is shaded in grey), while the waiting-time 
threshold $V_{\mathrm{th}}^{\mathrm{(wt)}}$ is used in the analysis of waiting times 
(e.g. internal and external waiting time, $T_{\mathrm{int}}$ and $T_{\mathrm{ext}}$). 
}
\label{Fig4}
\end{center}
\end{figure}

As a part of the overall response signal, the pure BN signal arises due to  the motion of the domain walls caused by 
flipping of magnetic moments tending to align with the local effective magnetic field. 
In ferromagnetic samples, this local field increases/decreases (almost everywhere) together with the external magnetic field, and 
therefore the corresponding values of induced electromotive force (EMF) are (almost always) one-sided. The domain wall motion proceeds in the 
form of one or several avalanches of finite duration possibly merging in time and space. Each such event induces a  BN pulse -  
 a sequence  of nonzero pure BN signal values of the same sign realized in contiguous short intervals of time.

Fig. \ref{Fig4} shows a part of the recorded signal suggesting that the 
response signal can be considered as a train of BN pulses, separated in time by the inactivity intervals of the sample  
during which the pure BN signal is absent. 
This means that the recorded signal can be decomposed into BN pulses, which 
is, however, not straightforward due to the superimposed induced voltage caused not by the rearrangements of the magnetic 
domains, but by other means. 
Indeed, some EMF is induced in the pickup coil because of the varying external magnetic 
field even without inserted sample. For the noiseless triangle time profile of the driving field, this contribution 
would appear in the signal time profile as a horizontal line switching its level at the half-period transitions, whereas in reality 
some concomitant noise,  caused by all factors except the change of sample's magnetization and therefore {\it external}, 
is inevitably 
superimposed as well. Due to this reason, several methods for 
baseline determination in the presence of noise (and investigated signal, here pure BN)  have been proposed so far, see e.g. 
\cite {SpcActa2005, A&A2008}. 
Here, like in \cite{Spasojevic1996}, we used the simplest and the fastest one
in which the baseline level $b_l$ is taken so that it corresponds to the discrete value of the digitized signal that is most 
frequently visited during the ongoing half-period, cf. Fig. \ref{Fig4}.

The next step in the decomposition of the recorded signal into BN pulses is the establishment of some threshold region around the 
baseline and subsequent recognition of the recorded signal parts lying outside that region as BN pulses. Here, 
the threshold region is taken as the range of signal values $v$ satisfying 
$b_l-V_{\mathrm{th}}<v<b_l+V_{\mathrm{th}}$ for some chosen threshold 
$V_{\mathrm{th}}>0$. As for the signal value $v$ close to $b_l$ one cannot resolve whether the external noise dominantly causes its part $v-b_l$ or not, the purpose of $V_{\mathrm{th}}$ is to discriminate between these two cases. So  for $|v-b_l|< V_{\mathrm{th}}$ 
(and small enough $V_{\mathrm{th}}>0$), one can consider that $v$ is dominantly caused by other (i.e. non-BN) causes and disregard 
such signal points from further analysis of BN distributions, whereas the remaining part of the recorded signal, despite being polluted by 
the external noise, is taken as the BN signal relative to the baseline and proceeded to further analysis. For the threshold $V_{\mathrm{th}}$ used in the foregoing way we say that it plays the role of a {\it base} threshold, in which case we will denote it as $V_{\mathrm{th}}^0$. 

The so-obtained BN signal is already decomposed into BN pulses, each being a subsequence 
$$\{v(t_s),v(t_s+\Delta t),...,v(t_e)\}\>$$ 
of the overall recorded sequence of digitized signal values $v(t)$
taken in the interval $\{t_s,t_s+\Delta t,...,t_e\}$ of contiguous discrete moments of acquisition time 
starting at $t_s$ and ending at $n\Delta t$ later moment $t_e$, where $\Delta t$ is the sampling interval 
($=5\times 10^{-5}$ s in our case), cf. Fig. \ref{Fig3}. 
$t_s$ and $t_e$ are defined in the usual way such that for a positive BN burst the signal goes above the threshold at $t=t_s$ and stays above the threshold until $t=t_e$, when it goes below the threshold for the first time after the start of the burst; note that for negative bursts, the opposite applies (i.e., the event starts at $t=t_s$, when the signal goes below the negative threshold and lasts until it goes above it at $t=t_s$).
Each BN pulse of a ferromagnetic sample is either positive or negative, i.e. all its values are either above $b_l+V_{\mathrm{th}}^0$ or below $b_l-V_{\mathrm{th}}^0$, depending on the sign of $dH/dt$, cf. Fig.~\ref{Fig2}.

The collection of BN pulses extracted via the foregoing procedure from the overall recorded signal depends on the choice of threshold $V_{\mathrm{th}}$. 
Because the basic role of imposing a threshold is to keep in further analysis most of the data points that are likely caused by 
BN reasons, a natural way would be to choose a threshold taking into account the width $w$ of the external noise. As explained 
in \cite{Spasojevic1996}, this width can be 
estimated from the data lying below the baseline and take $V_{\mathrm{th}}^0$ proportional to this 
width, $V_{\mathrm{th}}^0=d_lw$, using reasonable discrimination levels $d_l$ (e.g. between 0.5 and 3). Alternatively, the base threshold can be chosen differently, e.g. fixed to some constant value approximately matching the width of external noise, as we did in this paper.


Besides playing the role of a base threshold that is used in the decomposition of the response signal into BN pulses, variable threshold values are used in the analysis of waiting times (see the next subsection) in which case such threshold will be referred to as the waiting-time threshold $V_{\mathrm{th}}^{\mathrm{(wt)}}$. The logic here is to first define a pulse with one threshold ($V_{\mathrm{th}}^0$), and then a new threshold with another value  ($V_{\mathrm{th}}^{\mathrm{(wt)}}$) in order to be able to classify the waiting times into internal (due to breaking of the bursts into sub-avalanches) and external ones (those separating the "original" bursts)~\cite{JSTAT2009}.

\subsection{Experimental BN distributions}\label{ExpBNDistribss}

Each BN pulse 
$\{v(t_s),v(t_s+\Delta t),...,v(t_e)\}$
is characterized by several parameters. Mostly analyzed are its size 
$$S=\sum_{k=0}^{n}[v(t_s+k\Delta t)-b_l]\Delta t\>,$$ 
duration 
$$T=t_e-t_s\>,$$ i.e. the time interval between its ending moment $t_e$ and starting moment $t_s$, energy 
$$E=\sum_{k=0}^n [v(t_s+k\Delta t)-b_l]^2\Delta t\>,$$ 
and amplitude 
$$A=max\{v(t_s)-b_l,...,v(t_e)-b_l\}\>.$$
Regarding the question of to what extent are the so-obtained parameter values affected by the external noise, one can take 
that while the width of (zero-mean) noise is comparatively small and its values uncorrelated, it can not affect much the individual parameter 
values for moderate and large BN pulses, and likely has no effect on their statistics.

\begin{figure*}[hptb!]
\begin{center}
\includegraphics [width=\textwidth,trim=0cm 0.3cm 1.5cm 0cm,clip=true]{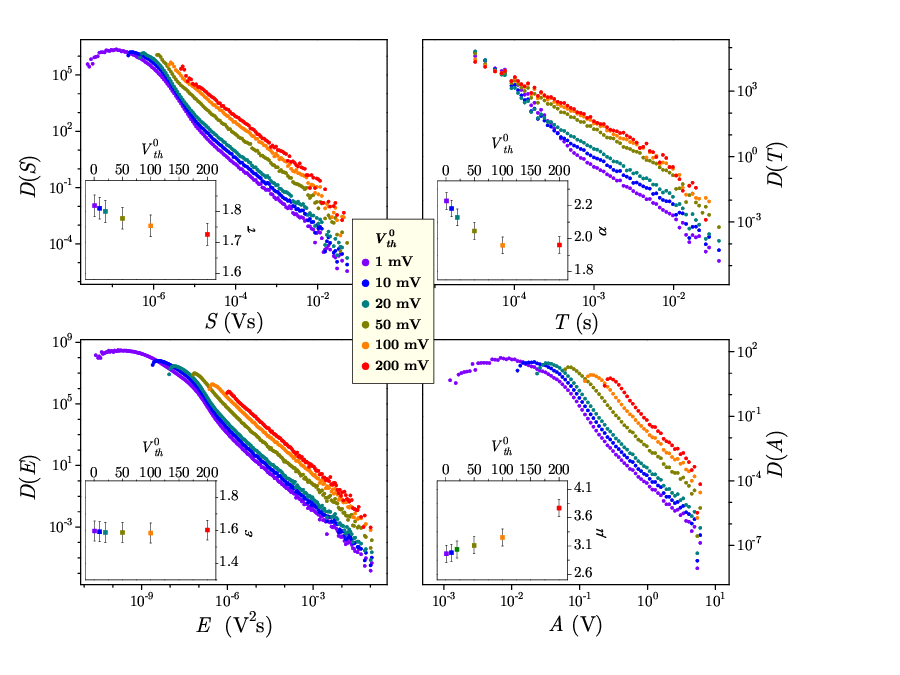}
\caption{
For the values of base threshold $V_{\mathrm{th}}^0$ quoted in the (common) legend, 
we give in main panels the integrated 
distributions of size $S$, duration $T$, 
energy $E$, and amplitude $A$ of BN pulses 
all collected from the 
data recorded in 20 cycles of the external magnetic field at the same driving rate $\Omega=5$~mHz. 
All distributions are normalized to the unit area at the lin-lin scale so as to be the experimental probability density functions. In the insets, we show the variation with $V_{\mathrm{th}}^0$ of the effective values of the distributions' exponents ($\tau$ for $D(S)$, $\alpha$  for $D(T)$, $\varepsilon$  for $D(E)$, and $\mu$ for $D(A)$) with the error bars mainly caused by the variation of the fitting region. 
}
\label{Fig5}
\end{center}
\end{figure*}

In Fig. \ref{Fig5} are presented the log-log plots of the experimental probability density functions (i.e. distributions with the unit area at the lin-lin scale) of size $S$, duration $T$, energy $E$, and amplitude $A$ of BN pulses extracted along the entire hysteresis loop at six values of base threshold $V_{\mathrm{th}}^0$ from the signals recorded at $5$~mHz. 
Previous studies of Barkhausen noise emissions (see e.g. \cite{Spasojevic1996,DUR06}) and our data from Fig. \ref{Fig5} show that these distributions follow the (modified) power-laws 
\begin{equation}
D_X(X)= \mathfrak {D}_X(X/X_0,X/X_1,...)/X^{a_X}\>,
\label{Eq:scaling}
\end{equation}
where $X$ stands for one of the avalanche 
parameters (size $S$, duration $T$, energy $E$, and amplitude $A$), and $a_X$ for the pertaining power-law exponent 
($\tau$, $\alpha$, 
$\varepsilon$ and $\mu$ for $X=S$, $T$, $E$, $A$, 
respectively). The exponents are 
associated with the slope of the log-log plot of 
$D_X(X)$ in its scaling region 
(i.e. the part in which
this plot appears as linear), while the cutoff function  
$\mathfrak {D}_X(X/X_0,X/X_1,...)$, depending on the cutoff parameters $X_0,X_1,...$, describes the departure of $D_X(X)$ from the power law shape at the distribution ends (e.g. $\mathfrak {D}_X(X/X_l,X/X_u)\approx const$ for $X_l\le X\le X_u$ for the cutoff function specified by the lower cutoff $X_l$ and the upper cutoff $X_u$).

The data displayed in Fig. \ref{Fig5} show that, although 
the choice of base threshold affects the shape of 
distributions, they all exhibit scaling regions which seem to be visually 
approximately parallel in their main part. This we quantified by extracting 
the effective values of the corresponding exponents (by 
the simple linear fit applied in the main part of the 
scaling region for each of them; see the comment 
\cite{FootNote1}) and presenting their variation with the 
chosen base threshold $V_{\mathrm{th}}^0$ in the inset 
for each of the distributions. Except for the energy 
exponent $\varepsilon$, the so-obtained exponent values 
are not constant, but instead show a systematic change 
with $V_{\mathrm{th}}^0$ maintaining (within the 
uncertainty bars) the fulfillment of the scaling 
relations \cite{Spasojevic1996,SpasojevicPRE2011,SpasojevicPRL2011}:
\begin{eqnarray}
\tau &=& 1+(\alpha-1)/\gamma_{S/T}\\
\varepsilon &=& 1+(\alpha-1)/(2\gamma_{S/T}-1)\\
\mu &=& 1+(\tau-1)/(1-1/\gamma_{S/T})
\label{ScalRel}
\end{eqnarray}
containing the exponent $\gamma_{S/T}$, see 
(\ref{s/t}), whose variation with $V_{\mathrm{th}}^0$ is shown in the left panel of Fig. \ref{Fig9}.

Regarding the choice of the base threshold values 
employed in Fig. \ref{Fig5}, we notice that, 
due to 5~mV noise width, the value of 1~mV base 
threshold is too small, enabling some signal parts 
dominated by the external noise to be recognized 
as (small) BN pulses, and also an occasional 
artificial merging of several 
separate BN pulses lined up in a sequence into larger 
ones. The 
next two values of $V_{\mathrm{th}}^0$, namely 10~mV and 
20~mV, should be the most appropriate at the 
current (i.e. 5~mV) noise width because they 
statistically eliminate the influence of external 
noise and do not discriminate recognition of  
BN pulses that are not too small. The distributions 
extracted at these two $V_{\mathrm{th}}^0$ values 
indicate the existence of two scaling regions like the 
ones observed in the case of the RFIM avalanche 
distributions of thin systems, see  
\cite{CrossoverPRE2018, PRE2019, BosaSciRep2019,StripPRE2020}. There it was shown 
that the small avalanches, not
reaching the system borders, propagate like
in bulky 3D systems, whereas the big
avalanches effectively propagate like 2D 
avalanches being sandwiched between the top 
and bottom system boundaries. 
So, the small 3D-like avalanches dominate 
in the initial steeper scaling region, 
followed by the less steep one caused by the big 
2D-like avalanches. The initial (i.e. 3D-like and 
steeper) part of the distributions' scaling 
region gradually vanishes with further 
increase of $V_{\mathrm{th}}^0$ and the distributions 
attain the shape of a single-slope power law 
sharply decreasing at the large avalanche end.

\begin{figure}[hptb!]
\begin{center}
\includegraphics[width=8.5cm,trim=0cm 2.8cm 2.2cm 0cm,clip=true ]{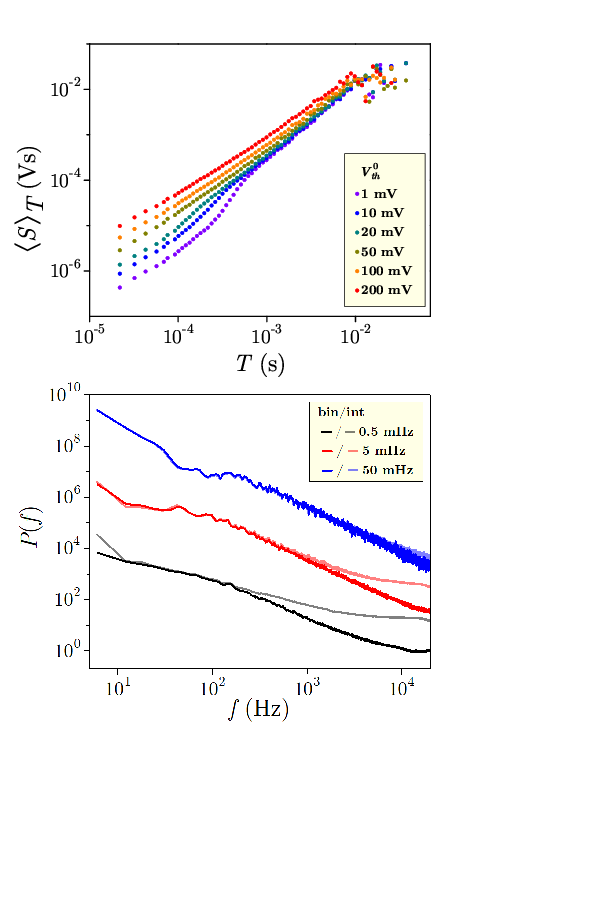}
\caption{Top panel: correlations between duration $T$ and the average size $\langle S\rangle _T$ of BN pulses with duration $T$ obtained for the same values of base threshold $V_{\mathrm{th}}^0$ as in Fig.~\ref{Fig5}. Bottom panel: BN power spectra $P(f)$ against frequency $f$ collected at three driving rates (0.5 mHz, 5 mHz, and 50 mHz) along the entire hysteresis loop (int) and in a narrow window (bin)  of the external magnetic field centered at the maximum of the average response signal. For both panels, the underlying sets of experimental data are the same as in Fig.~\ref{Fig5}.   
}
\label{Fig6}
\end{center}
\end{figure}

Experimental BN studies performed so far revealed that between the duration $T$ and the average size $\langle S\rangle_T$ of BN pulses of duration $T$ 
should exist the correlation of 
a power-law type
\begin{equation}
\langle S \rangle_T \sim T^{\gamma_{S/T}}\>,
\label{s/t}
\end{equation}
specified by the power-law exponent 
$\gamma_{S/T}$. Our 5~mHz data from the top panel of Fig. \ref{Fig6} suggest that the 
experimental value of this exponent, determined from the slope of the scaling region in its large-avalanche (linear) part, slightly depends on the choice of base 
threshold as is visible in the figure. Moreover, 
our data for $V_{\mathrm{th}}^0<50$~mV indicate the presence of two scaling parts in graphs. The initial one 
shows the correlations between $\langle S\rangle_T$ and $T$ for small avalanches and   continues through a transitional region to the main part of the scaling region giving the correlations for larger avalanches.

The BN power spectrum $P(f)$, i.e. the (spectral) density of power released by the BN signal at frequency $f$, is expected to follow the power law 
\begin{equation}
P(f)\sim f^{-\gamma_\mathrm{spc}}\>,
\label{spc}
\end{equation}
specified by the power exponent $\gamma_\mathrm{spc}$. In experiments, the BN signal is polluted by the external noise, so that the experimental power spectrum deviates from the power law most notably at the higher frequencies at which the noise dominates the Fourier components of the BN signal. This is noticeable in the  experimental power spectra, exemplified in the bottom panel of Fig. \ref{Fig6} by the integral spectra recorded at three driving rates (0.5~mHz, 5~mHz, and 50~mHz) along the entire hysteresis loop (and shown by the pale lines). The clearest example is the 0.5~mHz integral spectrum whose power-law part is absent due to the proportionally longest interval of time virtually without the BN noise. As the driving rate increases, the interval with a pronounced BN signal gets proportionally longer so does the power-law spectrum part. To reduce the influence of noise, we also calculated the binned power spectra, i.e. the power spectrum of the response signal recorded in a narrow window of the external magnetic field centered at that value of $H$ at which the averaged response signal attains its maximum. Under our experimental conditions, the binned spectra seem to be only weakly polluted by the external noise up $\sim 20$~kHz, so their power-law part extends that far.

As previously mentioned, thresholds are also used in 
the analysis of waiting times which are 
defined in the following way.
For any threshold 
$V_{\mathrm{th}}^{\mathrm{(wt)}}$,  chosen for 
the waiting-time analysis, 
some parts of the response signal remain 
below the imposed waiting-time threshold level 
$b_l+V_{\mathrm{th}}^{\mathrm{(wt)}}$, 
meaning that at any moment $t$ in such 
part $v(t)<b_l+V_{\mathrm{th}}^{\mathrm{(wt)}}$. 
The start $t_s$ and end $t_e$ of each of the 
corresponding intervals of time are 
(figuratively speaking) determined by two 
successive intersections of the response signal 
with the waiting-time threshold level, see in Fig.~\ref{Fig4}. So, one can take the duration $T_{\mathrm{w}}=t_e-t_s$ between these two moments $t_s$ and $t_e$  as the {\it waiting time}, and classify it either as the {\it external} waiting time $T_\mathrm{ext}$ or {\it internal} waiting time $T_\mathrm{int}$ if the moments $t_s$ and $t_e$ belong to different/same activity event, respectively. 

\begin{figure}[hptb!]
\begin{center}
\includegraphics[width=8.5cm,trim=0cm 1.5cm 0cm 0cm,clip=true ]{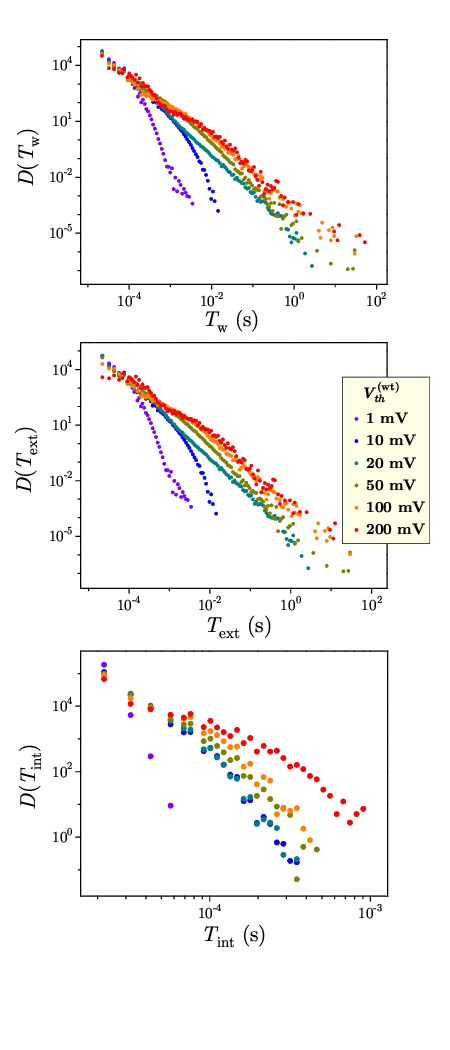}
\caption{Distributions of various types of waiting times: total  $T_{\mathrm{w}}$ in the top, external $T_{\mathrm{ext}}$ in the middle, and internal $T_{\mathrm{w}}$ in the bottom panel. The distributions are obtained at the 5~mHz driving frequency for the same threshold levels and the same sets of data as in Fig.~\ref{Fig5}.}
\label{Fig7}
\end{center}
\end{figure}

In Fig.~\ref{Fig7}, we 
present three types of distributions of waiting time: 
total waiting time $T_{\mathrm{w}}$ in the top panel, 
external waiting time  
$T_\mathrm{ext}$ in the middle panel, and
internal waiting time  
$T_\mathrm{int}$ in the bottom panel. 
These distributions become of a 
power-law type for the sufficiently high 
values of waiting-time threshold 
$V_{\mathrm{th}}^{\mathrm{(wt)}}$ signifying 
the presence of temporal correlations; 
otherwise, they are 
exponential, showcasing the random waiting 
times and absence of temporal correlations.

\subsection{Comparison with RFIM}\label{Comp}

In this section, we compare our experimental 
results with the results of 
numerical simulations of the athermal 
nonequilibrium Random Field Ising Model \cite{BelangerNatterman,Sethna2006} driven at finite rates with the aim to test its suitability for interpretation of the behavior of field-driven disordered ferromagnets. Since at room temperature conditions, like 
in our experiment, thermal fluctuations in real 
systems should not be important, we employed the athermal (i.e. zero-temperature) model version \cite{SethnaPRL93,OlgaPRB1999,Sethna2006}
which is significantly less demanding than 
the thermal (i.e., finite-temperature) version. In 
the athermal version the distribution of the 
random field is quenched and therefore the model 
behavior is fully deterministic, meaning that 
driving the system through the next hysteresis 
cycle gives the same results. This is not 
consistent with BN experiments having a somewhat 
different response in repeated hysteresis cycles that might originate, e.g., from thermal noise and/or redistribution of stress within the sample from cycle to cycle.
To overcome this deficiency, simulations are performed 
using a different configuration of  
the random field in each run, and the so obtained 
results are averaged over employed configurations 
having all the same value of the disorder 
parameter  (this is known as quenched averaging).

\subsubsection{On the model and numerical simulations}\label{Dtls}


In the Random Field Ising Model, 
the Ising spins $s_i=\pm1$, located at the 
sites $i$ of a lattice, interact with their 
nearest neighbors ferromagnetically, 
and are exposed to a time-varying homogeneous 
external magnetic field $H$ and a quenched 
random magnetic field taking uncorrelated  
values $h_i$ from a zero-centered Gaussian 
distribution with the standard deviation $R$  measuring disorder in the system.

The spin $s_i$ is stable at the moment $t_m$ of model time  if $h_i^{\mathrm{eff}}(t_m)s_i\ge 0$, where 
\begin{equation}
h_i^{\mathrm{eff}}(t_m)=\sum_{j}s_j^{(i)}+H(t_m)+h_i\>,
\label{Heff}
\end{equation}
 is the effective field acting on $s_i$ at the moment $t_m$ 
 encompassing the influence of its nearest neighbors $s_j^{(i)}$ and 
 the external field $H$ (both taken at that moment $t_m$), and the 
 random field $h_i$ at its site; 
otherwise, $s_i$ is unstable, and all unstable spins will flip  
reducing the value of the system Hamiltonian

\begin{equation}
\mathcal{H}=-\sum_{\langle ij\rangle} s_is_j-H\sum_i s_i-\sum_i h_is_i\>,
\label{Hamiltonian}
\end{equation}
at the next moment  $t_m+\Delta t_m$  of the (discrete) model time with $\Delta t_m=1$. 
Such spin-flipping spreads like 
an avalanche until all spins 
in the system become stable.

In the adiabatic regime \cite{Sethna2006,JSTAT21} each 
avalanche is nucleated due to such 
change in 
the external magnetic field $H$ that 
destabilizes only the least stable 
spin; thereafter, 
$H$ is kept constant as long as the 
nucleated avalanche lasts 
propagating over the shell of the 
nearest neighbors of 
spins flipped at the previous moment, 
the shell being placed at the rim of the cluster of spins flipped during the ongoing avalanche.

On the other hand, in the finite rate driving 
regime, the incessant change of the external 
magnetic field facilitates the propagation of the 
ongoing avalanche(s) and occasionally causes the 
nucleation of new ones
\cite{Tadic99,DahmenPRL2003,FDRJSTAT2021,Chaos2022}. 
This leads to the overlapping in time of such 
separately nucleated avalanches and possibly 
their merging 
in space forming a single activity event 
\cite{JSTAT_FDR_CORRS}. The activity events are separated in time by 
the intervals of the system's inactivity, 
leading in this model version to the natural choice of baseline 
level $b_l=0$ and identification of the event parameters 
(size $S$, duration $T$, energy $E$, and amplitude $A$) like in the 
experimental case, but after setting the value of base threshold in simulations to $V_{\mathrm{th}}^0=1$.

Because of the symmetry between the rising and falling part of the hysteresis loop, 
we performed the numerical RFIM simulations 
only along the rising part by the field $H(t)$ 
increasing at some driving rate $\Omega=\Delta H/\Delta t_m$ which was constant  like in our experiments. Each simulation starts with all 
spins set to -1 and some concordant big negative value of $H$,  
and afterward $H$ is  increased until all the spins are flipped to 1. 
We performed the simulations with the aid of 
the so-called sorted list algorithm (see in \cite{Kuntz1999,SpasojevicPRE2011}) modified and adjusted for the finite-rate driving regime.



\subsubsection{
Comparison of results of experimental BN measurements and numerical simulations of RFIM
}\label{CompBNandSim}

In order to optimally match the  simulation results to experimental, 
one can adjust in simulations the shape and size of the lattice, the 
value of disorder $R$, and the value of  driving rate $\Omega$. To 
this end, we performed simulations using the 
$32768$ x $2048$ x $8$ strip-like cubic  lattice with open boundaries mimicking our experimental samples by the ratio of its sides. 

Next, because of unknown value of disorder which might provide results similar to the experimental, we 
performed simulations for several values representing all three domains of disorder for the employed lattice (below critical, transitional, and above critical, see in \cite{JSTAT21}),
 and for each of them at a set of values of $\Omega$ from  $10^{-10}$ to  $10^{-6}$ covering for this lattice all types of 
driving regimes from slow (without spanning events \cite{RechePRB2003,RechePRB2004}) to fast (large spanning events of quasi-2D type \cite{SpasojevicPRE2014}), see in \cite{FDRJSTAT2021,JSTAT_FDR_CORRS}.

For each pair of simulation parameters 
$(R,\Omega )$ we performed 20 simulations with 
different realizations of the random magnetic 
field. As the activity events in simulations are 
clearly extraditable as the longest subsequences 
of the non-zero values appearing in the 
(noiseless) response signal at contiguous moments 
of time, the collecting of their statistics would 
be natural to perform with the base threshold 
$V_{\mathrm{th}}^0=1$. Still, for the intended 
comparison with experiments, we did this for all 
distributions at the 
same base threshold  
$V_{\mathrm{th}}^{\mathrm{sim}}=5$  in simulations 
and at the same base threshold 
in experiments $V_{\mathrm{th}}^{\mathrm{exp}}=50$~mV. Despite the 
$(V_{\mathrm{th}}^{\mathrm{exp}},V_{\mathrm{th}}^{\mathrm{sim}})$ pairs, chosen 
individually for each pair of the compared 
distributions, would provide closer matching, for 
simplicity we decided to perform all comparisons using the above pair of fixed thresholds with a remark 
that at and above 
$V_{\mathrm{th}}^{\mathrm{exp}}=50$~mV the 
$\langle S\rangle_T-T$ experimental correlations have the 
scaling region with a single slope, c.f. Fig. 
\ref{Fig6}, and that for  
$V_{\mathrm{th}}^{\mathrm{sim}}=5$ and all 
employed rates the ratio 
(RMS of simulation 
signal)/$V_{\mathrm{th}}^{\mathrm{sim}}$ is 
roughly the same as in experiment at $V_{\mathrm{th}}^{\mathrm{exp}}=50$~mV. 

\begin{figure*}[hptb!]
\begin{center}
\includegraphics[width=\textwidth,trim=0cm 2.2cm 1.5cm 0cm,clip=true] {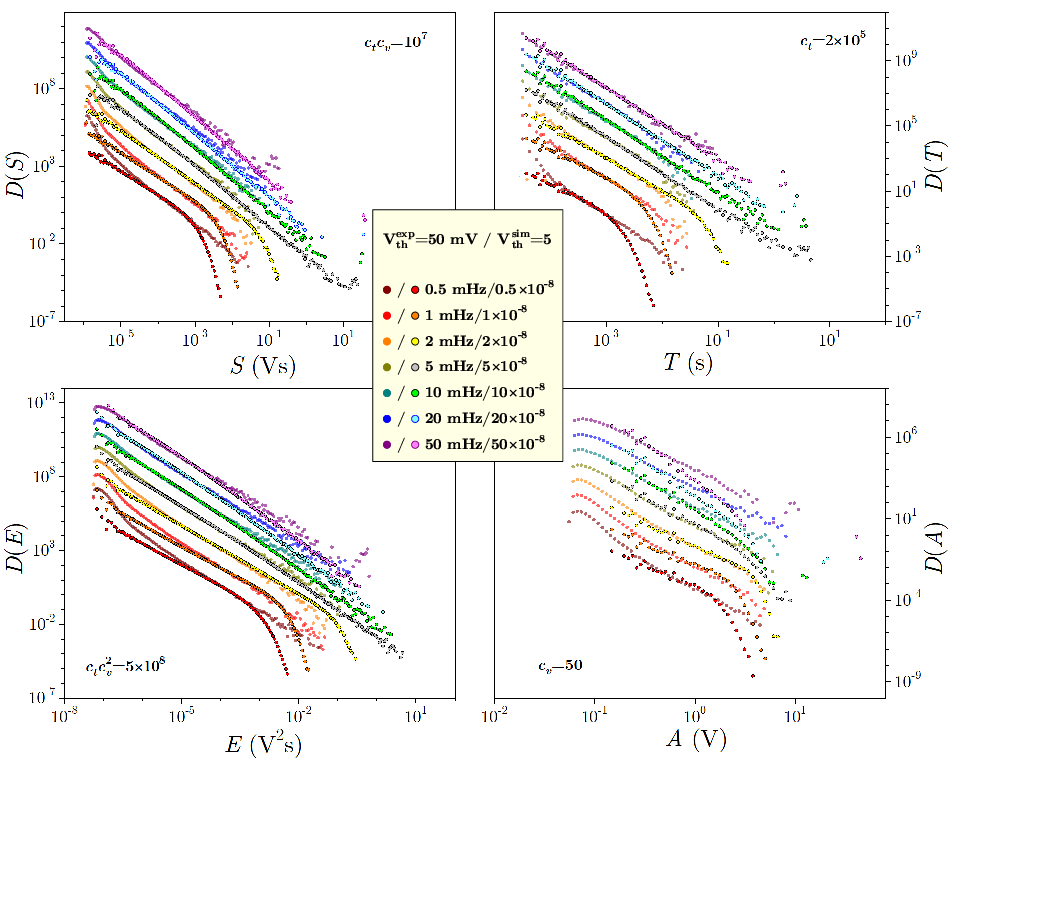}	
\caption{
Comparison of integrated (unit area) distributions of avalanche parameters (size $S$, duration $T$, energy $E$, and amplitude $A$) obtained in experiments and numerical RFIM simulations for the experimental/simulational driving rates quoted in the common legend. Each experimental distribution is extracted at the (same) experimental base threshold $V_{\mathrm{th}}^{\mathrm{exp}}=50$~mV out of 20 hysteresis cycles data and presented with full symbols on the requisite scale using SI units for time and voltage. Starting from the distribution recorded at the lowest driving rate, each experimental distribution obtained at the next (higher) rate is for better visibility vertically translated by one decade upwards relative to the distribution recorded at the previous (lower) rate. Each simulational distribution is extracted at the (same) simulational base threshold $V_{\mathrm{th}}^{\mathrm{sim}}=5$  out of 20 RFIM simulations on the $32768$x$2048$x$8$ cubic lattice performed with different realizations of the random magnetic field with disorder $R=2.3$. For comparison, the simulational distributions, presented by empty symbols, are shifted along the horizontal axis  dividing the data by a suitable factor ($c_t=2\times 10^5$ for $T$-axis, $c_v=50$ for $A$-axis, $c_tc_v=1\times 10^5$ for $S$-axis, and $c_tc_v^2=5\times 10^8$ for $E$-axis).}
\label{Fig8}
\end{center}
\end{figure*}

In Fig. \ref{Fig8} we compare the integrated 
distributions of avalanche event parameters  
(size, duration, energy and amplitude) collected 
in our experimental measurements and in numerical 
simulations. The presented simulational 
distributions are obtained for disorder $R=2.3$, 
being above the effective critical disorder for 
the adiabatically driven $32768$ x $2048$ x $8$ 
 RFIM system \cite{StripPRE2020}, and at a set of 
 "nice" values of the driving rate quoted in 
legend chosen in the same progression as 
the experimental ones. Among the tested driving rates and disorder values, ranging between $R=1.8$ and $R=3.0$ in 0.1 increments, this combination of driving rates and disorder provided the  
best achieved matching between experimental and simulational distributions presented in this and all subsequent figures. 

Because of the different time and signal scales in 
the experiment and in 
 simulations (real scales in the experiment, and 
 discrete model scales in simulations), the 
 matching of the two is  achieved by dividing the simulational time scale by the factor $c_t=2\times 10^5$ (equal to the 
 sampling rate used in our experiment) and the 
 signal scale by the factor $c_v=50$ (hence, the 
 factor $c_tc_v=10^7$ for the scale of avalanche 
 size and the factor $c_tc_v^2=5\times 10^8$ for 
 the scale of avalanche energy) providing the
 best matching with the experimental 
 distributions. In addition to this, due to the 
 existing difference in both the shape and the 
 logarithmic span, the distributions obtained 
 from the simulations are shifted along the 
 vertical axis so to attain the best overlapping 
 in the scaling regions of the pairing 
 distributions obtained in the experiment.

 The data presented in Fig. \ref{Fig8} show a significant overlapping for the pairs of distributions at faster rates and overlapping in the scaling region, but noticeable discrepancies in the initial and final part of the distributions at slower rates. The overlapping is the least in the case of amplitude distributions which could be related to a significantly smaller dynamic range of the simulated response signal at the chosen lattice size.

\begin{figure*}[hptb!]
\begin{center}
\includegraphics[width=\textwidth,trim=0cm 3.5cm 4cm 0cm,clip=true] {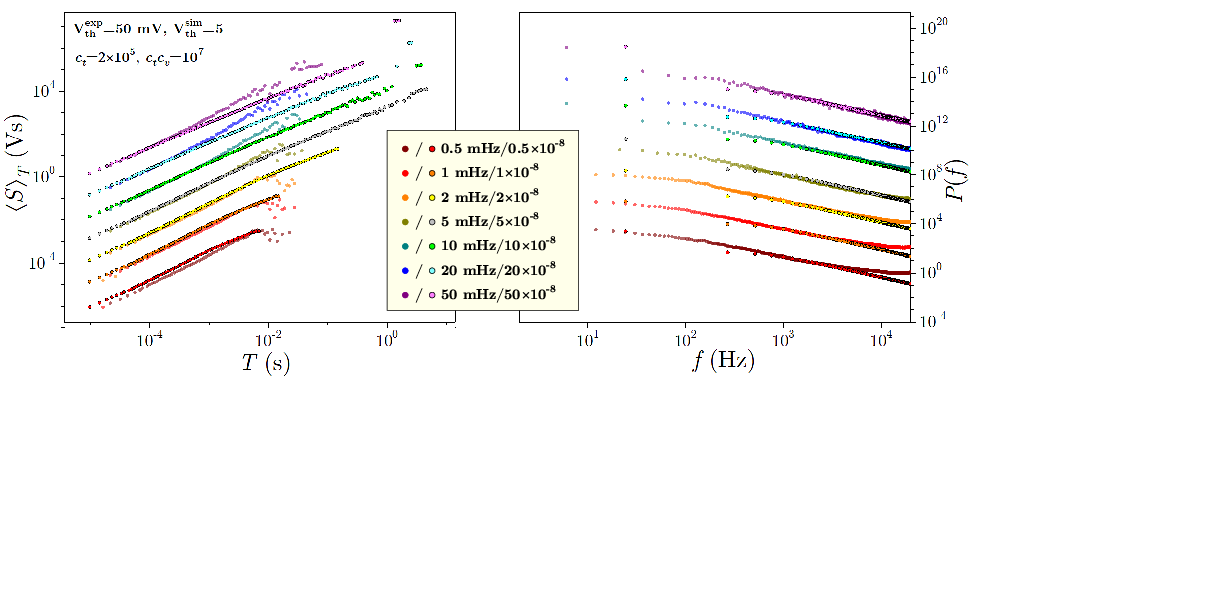} 
\caption{
Left panel: comparison of the experimental and simulational correlations between the avalanche duration $T$ and the average avalanche  size $\langle S\rangle_T$ of that duration extracted at the same experimental and simulational base thresholds, the same values of experimental and simulational driving rates, and the same $c_t$ and $c_tc_v$ factors as in Fig. \ref{Fig8}.
Right panel: comparison of the experimental and simulational power spectra $P(f)$ for the driving rates from the legend. Simulational frequencies are multiplied by the factor $c_t=2\times 10^{5}$. For visibility, each of the 
next-driving-rate curves in both panels is shifted vertically upwards by 
one/two decades in left/right panel relative to the previous one.  
The underlying sets of data are the same as in Fig. \ref{Fig8}. 
}
\label{Fig9}
\end{center}
\end{figure*}

\begin{figure*}[hptb!]
\begin{center}
\includegraphics[width=\textwidth,trim=0cm 1cm 2cm 0cm,clip=true] {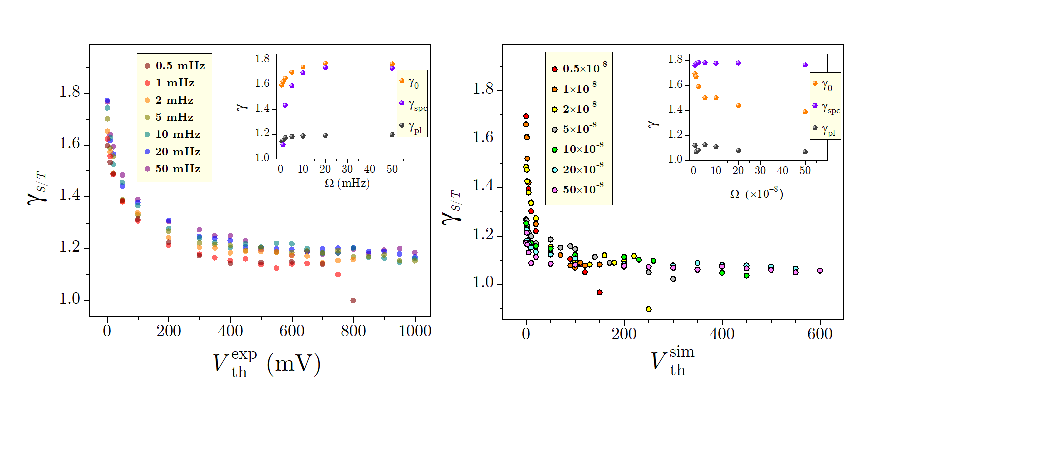}	
\caption{Left panel: effective experimental values of the exponent $\gamma_{S/T}$ against the base threshold $V_\mathrm{th}^\mathrm{exp}$. In the inset, we show against the driving rate $\Omega$ the effective experimental values of $\gamma_0$ (i.e. the value of $\gamma_{S/T}$ at the current driving rate $\Omega$ for the smallest experimental base threshold $V_\mathrm{th}^\mathrm{exp}=1$~mV), $\gamma_\mathrm{spc}$ (i.e. power spectrum exponent), and $\gamma_\mathrm{pl}$ (i.e. plateau value of the exponent $\gamma_{S/T}$ at the corresponding driving rate $\Omega$). Right panel: the same as in the left panel, but for the values obtained from the simulational data. Each effective exponent value is the slope  determined by the linear fit in the power law region of the corresponding distribution. The underlying data sets and other relevant parameters are the same as in Fig. \ref{Fig8}. 
}
\label{Fig10}
\end{center}
\end{figure*}

Comparison between the experimental and simulational  $\langle S\rangle_T-T$ correlations, presented in the left panel of Fig. \ref{Fig9}, is performed for the same sets of experimental and simulational data as in Fig. \ref{Fig8} using the same $c_t$ and $c_tc_v$ factors and the same threshold pairs. The best overlapping is obtained between $1$~mHz/$1\times 10^{-8}$ and $2$~mHz/$2\times 10^{-8}$ experimental/simulational driving rates, while for higher rates the experimental correlations depart more and more from the power law, likely due to spatial merging of avalanches occurring at these rates.

Even more intricate is the behavior of power spectra. Comparison of the experimental and simulational binned spectra for the used experimental/simulational driving rates is shown in the right panel of Fig. \ref{Fig9}. The compared data suggest that the 
experimental BN (binned) spectra can be described by the power law 
\begin{equation}
P(f)\sim f^{-\gamma_{S/T}}\>,
\label{spc}
\end{equation}
meaning that the power exponents $\gamma_\mathrm{spc}$ and $\gamma_{S/T}$ might be the same, as it was suggested in \cite{KuntzSpctrm} within the RFIM context. 
Here, however, one must take into account that 
both experimental and simulational effective 
values of the exponent $\gamma_{S/T}$ depend on 
the choice of the imposed base threshold 
$V_\mathrm{th}^0$ and on the driving rate $\Omega$, which we illustrated in 
Fig.~\ref{Fig10}. The data displayed in this 
figure show that both $\gamma_\mathrm{spc}$ and 
$\gamma_{S/T}$ vary with threshold $V_\mathrm{th}^0$ in a similar way decreasing 
from their maximum values at zero base threshold 
towards values at a plateau, one in the 
experimental and the other ($\sim 0.1$ lower) in the simulational
case, attained at rather high base thresholds. For each fixed value of base threshold $V_\mathrm{th}^0$, the variation of $\gamma_{S/T}$ with $\Omega$ is less, but present, including the plateau values. On the other hand, the effective values of $\gamma_\mathrm{spc}$ for different $\Omega$, displayed in the insets of these main panels, seem to be the same in the experimental case as the corresponding values of $\gamma_0$ (i.e. the effective value of $\gamma_{S/T}$ for the smallest base threshold $V_\mathrm{th}^0$ at the current driving rate $\Omega$), but not in the simulational case except for the very small driving rates. However, in both experimental and simulational cases, the power exponent $\gamma_\mathrm{spc}$ is undoubtedly different for the corresponding plateau values at all driving rates.

\begin{figure}[hptb!]
\begin{center}
\includegraphics[width=8.5cm,trim=0cm 0.5cm 1cm 0cm,clip=true]{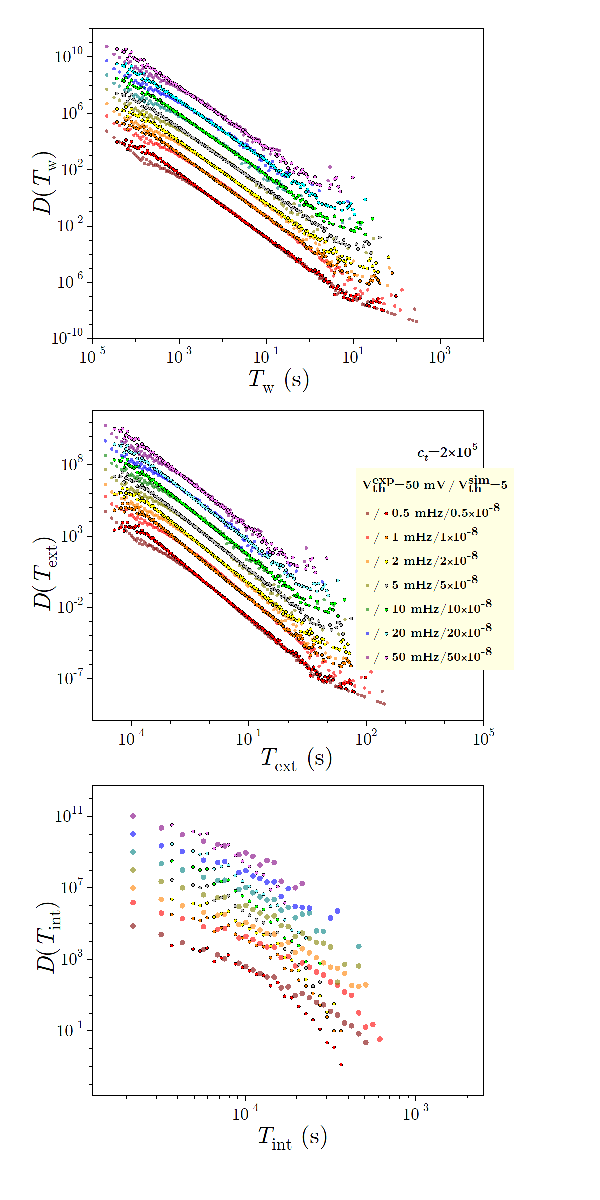}	
\caption{
Comparison of integrated distributions of total, external and internal waiting times, respectively, obtained in experiments and numerical simulations. The underlying data sets and all other parameters are the same as in Fig. \ref{Fig8}. 
}
\label{Fig11}
\end{center}
\end{figure}

In Fig. \ref{Fig11} we compare the distributions of waiting time: total, external, and internal shown in the top, middle, and bottom panel, respectively. The distributions calculated from the experimental and simulated data are overlapped to a high degree in the whole range of waiting times, except for the distributions of internal waiting times which overlap only in their scaling regions.

\begin{figure}[hptb!]
\begin{center}
\includegraphics[width=8.5cm,trim=0cm 0.5cm 0.5cm 0cm,clip=true]{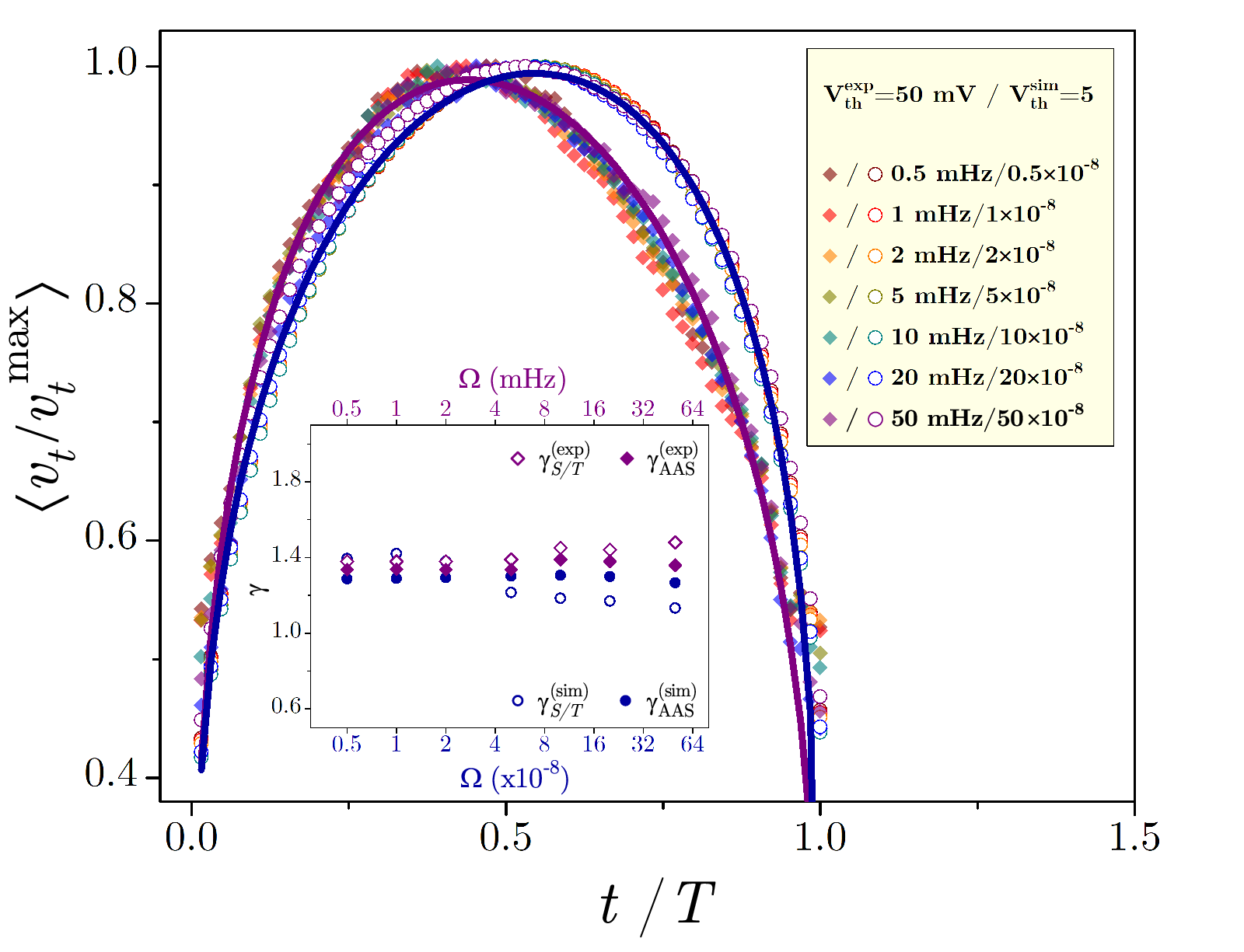}	
\caption{Average avalanche shapes,  shown by diamonds  obtained from the experimental and by circles obtained from the simulational data. The graph shows against $t/T$ (i.e. time $t$ measured from the start of avalanche scaled by avalanche duration $T$) the  values of $\langle v_t/ v_t^\mathrm{max}\rangle$ (i.e. the average value of the response signal $v_t$ scaled by its maximum value $v_t^\mathrm{max}$ during the avalanche).  
The underlying data sets and all other parameters are the same as in Fig. \ref{Fig8}. Full lines are fits to the functional form (\ref{AASfit}) and the inset shows the comparison of $\gamma_{S/T}$ to $\gamma_{AAS}$ values for both simulations and experiment.}
\label{fig12}
\end{center}
\end{figure}

The average avalanche shape \cite{shape} is given by $\langle v_t/v_t^\mathrm{max}\rangle$ (i.e. the average value of the response signal $v_t$ scaled by its maximum value $v_t^\mathrm{max}$ during the avalanche) taken as a function of $t/T$ (i.e. the time $t$ measured from the beginning of avalanche scaled by the avalanche duration $T$). Fig. \ref{fig12} shows that both in experiment and simulations these shapes are parabolic-like and rather similar, the experimental being slightly right-skewed and simulational (even less) left-skewed. Performing the fits of the average avalanche shape data to the functional form (found in \cite{shape})
\begin{equation}
\langle V(t|T) \rangle \propto T^{\gamma-1}\Big[\frac{t}{T}\Big(1-\frac{t}{T}\Big)\Big]^{\gamma-1}\Big[1-a\Big(\frac{t}{T}-\frac{1}{2}\Big)\Big]
\label{AASfit}
\end{equation}
allows us to estimate exponent $\gamma$ as the fitting parameter;
here parameter $a$ accounts for the underlying asymmetry of the average avalanche shape. 
So-estimated values of $\gamma$ are contrasted to the $\gamma_{S/T}$ data and are shown in the inset of Fig.\ref{fig12} against the driving rate $\Omega$. One can observe that the values remain fairly close and consistent in the region of low driving rates, but begin to vary more and more as the driving rate increases. Two of the representative fits to the function (\ref{AASfit}) are included in the main panel of Fig. \ref{fig12}, shown with full line.

\begin{figure*}[hptb!]
\begin{center}
\includegraphics[width=\textwidth,trim=0cm 2.2cm 2cm 0cm,clip=true] {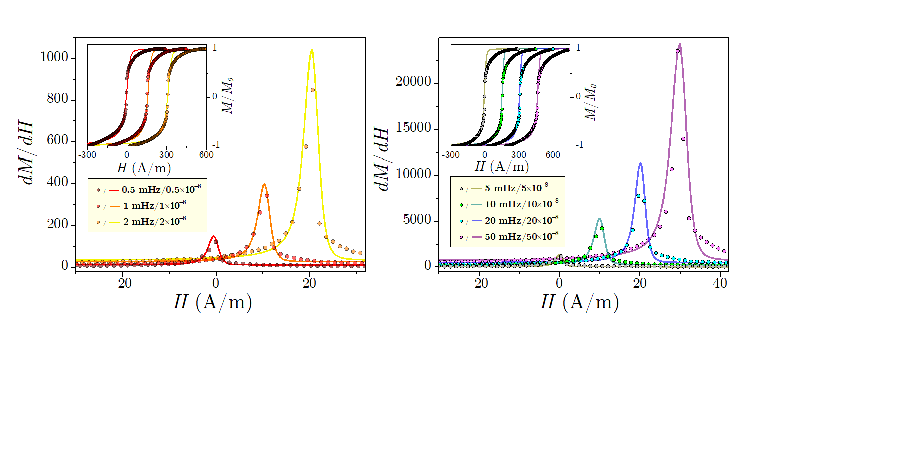}	
\caption{
Comparison of averaged magnetization and magnetic susceptibility curves obtained in experiments (symbols) and numerical simulations (full lines), the later scaled so as to achieve a matching with the experimental. The underlying data sets and other relevant parameters are the same as in Fig. \ref{Fig8}. In the main panels, we show the magnetic susceptibilities $dM/dH$ against the external magnetic field $H$, while in the insets we show  $M/M_0$ (i.e. the magnetization $M$ scaled by the saturation magnetization $M_0$) against $H$; both types of curves correspond to the rising part of the hysteresis loop, and the $dM/dH$ experimental curves are inverted (i.e. multiplied by -1), c.f. Fig.~\ref{Fig2}. In main panels/insets the curves are shifted for better visibility, namely susceptibilities/magnetizations in increments of 10~Am${}^{-1}$/150~Am${}^{-1}$ relative to the (unshifted) curves which correspond to the 0.5~mHz/5~mHz driving rates in left/right panel.   
}
\label{Fig13}
\end{center}
\end{figure*}

Finally, in Fig. \ref{Fig13} we contrast the magnetizations and magnetic susceptibilities obtained from the experiment and the numerical simulations. Magnetizations, rescaled by the saturation value $M_0$ and susceptibilities $dM/dH$, are presented against the external magnetic field $H$. Regarding the behavior of these curves below and above the coercive field $H_c$ 
(i.e. the value of the external field $H$ at which $M=0$),  one can see that 
the matching of data obtained from the experiments and numerical simulations is very good for $H<H_c$ and notably worse for $H>H_c$.

\section{Discussion and Conclusion}
\label{concl}

The main issue we faced in the comparison between the results obtained in our BN experiment and the possible results of the nonequilibrium athermal RFIM numerical simulations was the identification of the appropriate values of the RFIM parameters (lattice sizes, disorder $R$, and driving rate $\Omega$) that would provide a reasonable matching of the two types of data.

To this end, we performed the 
simulations on the $32768\times 2048\times 8$ 
strip-like cubic lattice, the largest one we 
could use. This lattice with more than half a 
billion spins has approximately the same 
aspect ratio as the sample used in our experiment, 
which is important because the simulational 
results are affected by the lattice aspect ratio 
and also by the lattice finite size \cite{CrossoverPRE2018,PRE2019,StripPRE2020,FDRJSTAT2021}. These two facts prevented us to use a 
bigger lattice (e.g. doubled along each of its 
sides) because seeking the reasonable 
$(R,\Omega)$ pairs would be out of our reach due 
to the greatly increased running time of 
simulations at the corresponding computer memory 
demands.

Regarding the optimality of the value of disorder 
and the set of values of the driving rate in 
simulations for which  the comparison with 
experimental findings is performed, we point out 
that these values depend on the choice of lattice 
(i.e. they might be somewhat different for 
another lattice due to the conformity of the 
finite-size and driving-rate scaling conditions \cite{FDRJSTAT2021}). Furthermore, even for the current choice of lattice, we do not claim that these values are really optimal, but merely such that provide a reasonable matching of the experimental and (scaled) simulational data. This is in particular because the employed version of the  RFIM is not indeed a realistic model for the Barkhausen noise which, instead of modeling the time evolution of magnetic domain pattern of driven disordered ferromagnetic samples responsible for the emission of BN, treats individual spins with the simplest mutual exchange coupling and the coupling with the quenched random magnetic field whose existence and characterization in real samples is still not well documented and understood. Nevertheless, as the employed RFIM version proved to be a paradigm for the description of the evolution of the nonequilibrium complex systems, we consider as valuable the comparison between its predictions and the characterization of the evolution  recorded in the Barkhausen noise measurements.

To conclude, in this paper we compared the 
findings of the Barkhausen noise measurements and 
the simulations of the nonequilibrium athermal 
Random Field Ising Model (RFIM). The measurements 
were performed on a VITROPERM~800 metallic glass 
strip driven by the external magnetic field at 
the rates between 0.5~mHz and 50~mHz. All RFIM 
simulations were accomplished using  suitable 
model parameters (lattice sizes with the aspect 
ratio as for the sample, a single value of 
disorder of the quenched random field, and a 
two-decade-wide range of driving). Applying these adjustments in simulations, allowed us to achieve 
a considerable matching with the experimental 
 data. We hope that our 
findings will be helpful in the interpretation 
and analysis of experimental results obtained 
from a variety of strip-like
disordered ferromagnetic samples driven at  
constant rates. Further on, the results of our 
study may also stimulate some future theoretical 
research, invoking the development of models that 
will more accurately capture the essence of the 
BN underlying dynamics.


\begin{acknowledgments}
We acknowledge the support by the Ministry of Science, Technological Development and Innovation of Republic of Serbia (Agreements 
No. 451-03-47/2023-01/200162 and  451-03-47/2023-01/200122). LL acknowledges the support of the Research Council of Finland via the Academy Project BarFume (Project no. 338955). 
AD acknowledges
the support of the Serbian Academy of Sciences and Arts via the Grant
F133. 
We thank Dr. Tatjana Sre\'ckovi\'c  
from the Institute for Multidisciplinary Research, University of Belgrade, Serbia, for the thermal treatment of our 
samples, and Viktor \'Cosi\'c from Custom Electronics d.o.o., Belgrade, Serbia, for his assistance in overcoming certain instruments malfunctions. \\
\end{acknowledgments}


\section*{References}
\begin{thebibliography}{99}
\bibitem{CracklingNoise}%
J. P. Sethna, K. A. Dahmen, and C. R. Myers, Nature {\bf 410}, 242-250 (2001)
\bibitem{Barkhausen}%
H. Barkhausen, Zwei mit Hilfe der neuen Verstarker entdeckte Erscheinugen, Physik Z. {\bf 20}, 401 (1919)
\bibitem{ABBM}%
B. Alessandro, C. Beatrice, G. Bertotti, and A. Montorsi, Domain-wall dynamics and Barkhausen effect in metallic ferromagnetic materials. II. Experiments, J. Appl. Phys. \textbf{68}, 2908–2915 (1990).
\bibitem {Spasojevic1996}%
D. Spasojevi\'c, S. Bukvi\'c, S. Milo\v{s}evi\'c and H. E. Stanley, Barkhausen noise: {E}lementary signals, power laws, and scaling relations, Phys. Rev. E \textbf{54}, 2531 (1996)
\bibitem{Bahiana}%
M. Bahiana, B. Koiller, S. L. A. de Queiroz, J. C. Denardin, and R. L. Sommer, Domain size effects in Barkhausen noise, Phys. Rev. \textbf{59}, 3884–3887 (1999)
\bibitem{Mehta}%
A. P. Mehta, A. C. Mills, K. A. Dahmen, and J. P. Sethna, Universal pulse shape scaling function and exponents: Critical test for
avalanche models applied to Barkhausen noise, Phys. Rev. E \textbf{65}, 046139 (2002)
\bibitem{Colaiori}%
F. Colaiori, S. Zapperi, and G. Durin, Shape of a Barkhausen pulse, J. Magn. Magn. Mater. \textbf{272–276}, E533–E534 (2004)
\bibitem{DZ2000}%
G. Durin, and S. Zapperi, Scaling Exponents for Barkhausen Avalanches in Polycrystalline and Amorphous Ferromagnets, Phys. Rev. Lett. \textbf{84}, 4705–4708 (2000)
\bibitem{Puppin}%
E. Puppin, Statistical Properties of Barkhausen Noise in Thin Fe Films, Phys. Rev. Lett. \textbf{84}, 5415–5418 (2000)
\bibitem{Moore}%
T. A. Moore, J. Rothman, Y. B. Xu, and J. A. C. Bland, Thickness-dependent dynamic hysteresis scaling behavior in epitaxial Fe/GaAs(001)
and Fe/InAs(001) ultrathin films, Jour. of App. Phys. \textbf{89} 7018 (2001)
\bibitem{Puppin2}%
E. Puppin, E. Pinotti, and M. Brenna, Barkhausen noise in variable thickness amorphous finemet films, J. Appl. Phys. \textbf{101}, 063903 (2007)
\bibitem{Kim}%
D.-H. Kim, S.-B. Choe, and S.-C. Shin, Direct Observation of Barkhausen Avalanche in Co Thin Films, Phys. Rev. Lett. \textbf{90}, 087203 (2003)
\bibitem{Ryu}%
K.-S. Ryu, H. Akinaga, and S.-C. Shin, Tunable scaling behaviour observed in Barkhausen criticality of a ferromagnetic film, Nat. Phys. \textbf{3}, 547–550 (2007)
\bibitem{Shin}%
S.-C. Shin, K.-S. Ryu, D.-H. Kim, S.-B. Choe, and H. Akinaga, Power-law scaling behavior in Barkhausen avalanches of ferromagnetic thin films, J. Magn. Magn. Mater. \textbf{310}, 2599–2603 (2007)
\bibitem{Lee}%
H.-S. Lee, K.-S. Ryu, K.-R. Jeon, S. S. P. Parkin, and S.-C. Shin, Breakdown of Barkhausen critical-scaling behavior with increasing domain-wall pinning in ferromagnetic films, Phys. Rev. B \textbf{83}, 060410 (2011)
\bibitem{Bohn1}%
F. Bohn, et al., Universal properties of magnetization dynamics in polycrystalline ferromagnetic films, Phys. Rev. E \textbf{88}, 032811 (2013)
\bibitem{Bohn2}%
F. Bohn, et al., Statistical properties of Barkhausen noise in amorphous ferromagnetic films, Phys. Rev. E \textbf{90}, 032821 (2014)
\bibitem{DurPRL2016}%
G. Durin, et al., Quantitative Scaling of Magnetic Avalanches, Phys. Rev. Lett. \textbf{117}, 087201 (2016)
\bibitem{Bohn3}%
F. Bohn, G. Durin, M. A. Correa, et al, Playing with universality classes of Barkhausen avalanches, Sci. Rep. \textbf{8}, 11294 (2018)
\bibitem{LasseTEM}%
M. Honkanen, S. Santa-aho, L. Laurson, N. Eslahi, A. Foi, M. Vippola,
Mimicking Barkhausen noise measurement by in-situ transmission electron microscopy - effect of microstructural steel features on Barkhausen noise, Acta Materialia \textbf{221}, 117378 (2021)
\bibitem{nano}%
J. Uhl, S. Pathak, D. Schorlemmer et al., Universal Quake Statistics: From Compressed Nanocrystals to Earthquakes, Sci. Rep. \textbf{5}, 16493 (2015)
\bibitem{imbib}%
S. Santucci et al., Avalanches of imbibition fronts: Towards critical pinning,  Europhys. Lett. \textbf{94}, 46005 (2011)
\bibitem{plastic}%
D. M. Dimiduk, C. Woodward, R. LeSar, and M. D. Uchic, Scale-Free Intermittent Flow
in Crystal Plasticity, Science \textbf{312}, 1188 (2006)
\bibitem{Ispanovity2014}
P. D. Ispanovity, L. Laurson, M. Zaiser, I. Groma, S. Zapperi, and M. J. Alava Avalanches in 2D Dislocation Systems: Plastic Yielding Is Not Depinning, Phys. Rev. Lett. {\bf 112}, 235501 (2014).
\bibitem{Jstat2015}%
S. Jani\'cevi\'c, M. Ovaska, M. J. Alava, and L. Laurson, Avalanches in 2D dislocation systems without applied stresses, J. Stat. Mech. P07016, (2015)
\bibitem{Ispanovity2022}%
P. D. Isp\'anovity, D. Ugi, G. P\'eterffy, et al. Dislocation avalanches are like earthquakes on the micron scale, Nat. Commun. \textbf{13}, 1975 (2022)
\bibitem{heart}%
P. Ch. Ivanov et al., From $1/f$ noise to multifractal cascades in heartbeat dynamics, Chaos: Interdisc. J. Nonlin. Sci. \textbf{11}, 641 (2001)
\bibitem{brain}%
C.-C. Lo et al., Dynamics of sleep-wake transitions during sleep, Europhys. Lett. \textbf{57}, 625 (2002)
\bibitem{DahmenPRL2012}%
N. Friedman, S. Ito, B. A. W. Brinkman, M. Shimono, R. E. L. DeVille, K. A. Dahmen, J. M. Beggs, and T. C. Butler,
Universal Critical Dynamics in High Resolution Neuronal Avalanche Data, Phys. Rev. Lett. \textbf{108}, 208102 (2012)
\bibitem{MillerSciRep2019}%
S. R. Miller, S. Yu, and D. Plenz,  The scale-invariant, temporal profile of neuronal avalanches in relation to cortical $\gamma$–oscillations, Sci Rep \textbf{9}, 16403 (2019); see also \emph{Criticality in Neural Systems}, edited by D. Plenz and E. Niebur, Wiley Online Library (2014)
\bibitem{FISH98}%
D. S. Fisher, Collective transport in random media: from superconductors to earthquakes, Phys. Rep. \textbf{301}, 113 (1998)
\bibitem{Earthquakes}%
J. Davidsen and M. Baiesi, Self-similar aftershock rates, Phys. Rev. E  \textbf{94}, 022314 (2016)
\bibitem{Petri}%
A. Bizzarri, A. Petri, A. Baldassarri, Earthquake dynamics constrained from laboratory experiments: new insights from granular materials, Ann. Geophys. \textbf{64}(4), SE 441 (2021)
\bibitem{financial}%
J. Perelló, J. Masoliver, A. Kasprzak, and R. Kutner, Model for interevent times with long tails and multifractality in human communications: An application to financial trading, Phys. Rev. E \textbf{78}, 036108 (2008)
\bibitem{Financial2013}%
J. P. Bouchaud, Crises and Collective Socio-Economic Phenomena: Simple Models and Challenges, J. Stat. Phys. \textbf{151}, 567-606 (2013)
\bibitem{BelangerNatterman}%
D. P. Belanger and T. Nattermann, \emph{Spin Glasses and Random Fields} edited by Young A P World Scientific Singapore, pp. 251--298 (1998)
\bibitem{SethnaPRL93}%
J. P. Sethna, K. A. Dahmen, S. Kartha, J. A. Krumhansl, B. W. Roberts and J. D. Shore, Hysteresis and hierarchies: {D}ynamics of disorder-driven first-order phase transformations, Phys. Rev. Lett. \textbf{70}, 3347 (1993)
\bibitem{OlgaPRB1999}%
O. Perkovi\'c, K. A. Dahmen and J. P. Sethna, Disorder-induced critical phenomena in hysteresis: {N}umerical scaling in three and higher dimensions, Phys. Rev. B \textbf{59}, 6106 (1999)
\bibitem{Sethna2006}%
J. P. Sethna, K. A. Dahmen and O. Perkovi\'c,  Random-field Ising models of hysteresis in \emph{The Science of Hysteresis} edited by Bertotti G and Mayergoyz I Academic Press Amsterdam Vol. \textbf{2}, pp. 107-179 (2006)
\bibitem{herranen}%
T. Herranen and L. Laurson, Barkhausen noise from precessional domain wall motion, Phys. Rev. Lett. {\bf 122}, 117205 (2019)
\bibitem{kaappa}%
S. Kaappa and L. Laurson, Barkhausen noise from formation of 360$^{\circ}$ domain walls in disordered permalloy thin films, Phys. Rev. Research {\bf 5}, L022006 (2023). 
\bibitem{stanley}%
S. Zapperi, P. Cizeau, G. Durin, and H. E. Stanley, Dynamics of a ferromagnetic domain wall: Avalanches, depinning transition, and the Barkhausen effect, Phys. Rev. B {\bf 58}, 6353 (1998).
\bibitem{FontViv}%
C. Frontera and E. Vives, Studying avalanches in the ground state of the two-dimensional random-field Ising model driven by an external field,
Phys. Rev. E \textbf{62}, 7470 (2000)
\bibitem{LiuDah}%
Y. Liu and K. A. Dahmen, Unexpected universality in static and
dynamic avalanches, Phys. Rev. E \textbf{79}, 061124 (2009)
\bibitem{LiuDah2}%
Y. Liu and K. A. Dahmen, Random-field Ising model in and out of equilibrium, Europhys. Lett. \textbf{86}, 56003 (2009)
\bibitem{Balog}%
I. Balog, G. Tarjus, and M. Tissier, Criticality of the random
field Ising model in and out of equilibrium: A nonperturbative
functional renormalization group description, Phys. Rev. B \textbf{97},
094204 (2018)
\bibitem{SpcActa2005} %
S. Bukvi\'c and Dj. Spasojevi\'c, 
An alternative approach to spectrum baseline estimation, 
Spectrochimica Acta Part B: Atomic Spectroscopy \textbf{60}, 1308 (2005)
\bibitem{A&A2008} %
S. Bukvi\'c, Dj. Spasojevi\'c, and V. \v Zigman
Advanced fit technique for astrophysical spectra, 
Astronomy \& Astrophysics \textbf{477}, 967 (2008)
\bibitem{JSTAT2009}%
L. Laurson, X. Illa, and M.J. Alava, The effect of thresholding on temporal avalanche statistics, J. Stat. Mech.: Theo. Exp., P01019 (2009)
\bibitem{DUR06}%
G. Durin G and S. Zapperi 2006  {\it The Science of Hysteresis}, edited by Bertotti, G. $\&$ Mayergoyz, I., Academic Press, Amsterdam \textbf{2} pp.181--267
\bibitem{FootNote1}
It is known that the exponent value extracted through the linear fit applied in the scaling region can be significantly influenced by the presence of the cutoffs. More reliable exponent values are obtained by fitting the data to a suitably chosen model function reasonably describing the cutoffs,  see e.g.  in \cite{Rosso} the analytic form (15) describing the cutoff function. 
\bibitem{SpasojevicPRE2011}%
D. Spasojevi\'{c}, S. Jani\'{c}evi\'{c} and M. Kne\v{z}evi\'{c}, Avalanche distributions in the two-dimensional nonequilibrium zero-temperature random field Ising model, Phys. Rev. E \textbf{84}, 051119 (2011)
\bibitem{SpasojevicPRL2011}%
D. Spasojevi\'{c}, S. Jani\'{c}evi\'{c} and M. Kne\v{z}evi\'{c}, Numerical {E}vidence for {C}ritical {B}ehavior of the {T}wo-{D}imensional {N}onequilibrium {Z}ero-{T}emperature {R}andom {F}ield {I}sing {M}odel, Phys. Rev. Lett. \textbf{106} 175701 (2011)
\bibitem{CrossoverPRE2018}%
D. Spasojevi\'{c}, S. Mijatovi\'{c}, V. Navas-Portela and E. Vives, Crossover from three-dimensional to two-dimensional systems in the nonequilibrium zero-temperature random-field Ising model, Phys. Rev. E \textbf{97}, 012109 (2018)
\bibitem{PRE2019}%
S. Mijatovi\'{c}, D. Jovkovi\'{c}, S. Jani\'{c}evi\'{c}, and D. Spasojevi\'{c}, Critical disorder and critical magnetic field of the nonequilibrium athermal random-field Ising model in thin systems, Phys. Rev. E \textbf{100}, 032113 (2019)
\bibitem{BosaSciRep2019}%
B. Tadi\'c, S. Mijatovi\'c, S. Jani\'cevi\'c, Dj. Spasojevi\'c and G. J. Rodgers, The critical {B}arkhausen avalanches in thin random-field ferromagnets with an open boundary,  Sci. Rep. \textbf{9} 6349 (2019)
\bibitem{StripPRE2020}%
S. Mijatovi\'{c}, M. Brankovi\'{c}, S. Graovac and D. Spasojevi\'{c}, Avalanche properties in striplike ferromagnetic systems, Phys. Rev. E \textbf{102}, 022124 (2020)
\bibitem{JSTAT21}%
S. Jani\'{c}evi\'{c}, D. Kne\v{z}evi\'{c}, S. Mijatovi\'{c} and D. Spasojevi\'{c}, Scaling domains in the nonequilibrium athermal random field {I}sing model of finite systems, J. Stat. Mech., 013202 (2021)
\bibitem{Tadic99}%
B. Tadi\'c, Dynamic criticality in driven disordered systems: role of depinning and driving rate in {B}arkhausen noise, Physica A \textbf{270}, 125-134 (1999)
\bibitem{DahmenPRL2003}%
R. A. White and K. A. Dahmen, Driving {R}ate {E}ffects on {C}rackling {N}oise, Phys. Rev. Lett. \textbf{91}, 085702 (2003)
\bibitem{FDRJSTAT2021}%
S. Radi{\'{c}}, S. Jani{\'{c}}evi{\'{c}}, D. Jovkovi{\'{c}} and D. Spasojevi{\'{c}}, The effect of finite driving rate on avalanche distributions, J. Stat. Mech. \textbf{2021} 093301 (2021)
\bibitem{Chaos2022}%
D. Spasojevi{\'{c}}, and S. Jani{\'{c}}evi{\'{c}}, Two-dimensional ferromagnetic systems with finite driving, Chaos, Solitons and Fractals \textbf{158}, 112033 (2022)
\bibitem{JSTAT_FDR_CORRS}%
D. Spasojevi\'{c}, S. Radi{\'{c}}, D. Jovkovi{\'{c}} and S. 
Jani\'{c}evi\'{c}, Spin activity correlations in driven
disordered systems,
Journal of Stat.Mech.: Theory and Experiment, 063302 (2022)
\bibitem{Kuntz1999}%
M. Kuntz, O. Perkovi\'{c}, K. A. Dahmen, B. W. Roberts and J. P. Sethna, Hysteresis, avalanches, and noise, Computing in Science Engineering \textbf{1}, 73 (1999)
\bibitem{RechePRB2003}%
F. J. P\'erez-Reche and E. Vives, Finite-size scaling analysis of the avalanches in the three-dimensional {G}aussian random-field Ising model with metastable dynamics, Phys. Rev. B \textbf{67}, 134421 (2003)
\bibitem{RechePRB2004}%
F. J. P\'erez-Reche and E. Vives, Spanning avalanches in the three-dimensional {G}aussian random-field {I}sing model with metastable dynamics: {F}ield dependence and geometrical properties, Phys. Rev. B \textbf{70}, 214422 (2004)
\bibitem{SpasojevicPRE2014}%
D. Spasojevi\'{c}, S. Jani\'{c}evi\'{c} and M. Kne\v{z}evi\'{c}, Analysis of spanning avalanches in the two-dimensional nonequilibrium zero-temperature random-field {I}sing model, Phys. Rev. E \textbf{89}, 012118 (2014)
\bibitem{KuntzSpctrm}%
M. C. Kuntz and J. P. Sethna, Noise in disordered systems: {T}he power spectrum and dynamic exponents in avalanche models, Phys. Rev. B \textbf{62}, 11699--11708 (2000)
\bibitem{Rosso}%
A. Rosso, P. Le Doussal, and K. J. Wiese,  Avalanche-size distribution at the depinning transition: A numerical test of the theory, Phys. Rev. B {\bf 80}, 144204 (2009)
\bibitem{shape}%
L. Laurson, X. Illa, S. Santucci, K. T. Tallakstad, K. J. Måløy, and M. J. Alava, Evolution of the average avalanche shape with the universality class, Nat. Commun. {\bf 4}, 2927 (2013)

\end{thebibliography}
\end{document}